\documentclass[aps,pre,floatfix,twocolumn]{revtex4} 
\usepackage{amsfonts}
\usepackage{epsf}
\usepackage{subfigure}
\usepackage{amsmath}	
\usepackage{amssymb}
\usepackage{graphicx}
\begin{document}
\title{Noise-induced escape from bifurcating attractors: \\ Symplectic approach in the weak-noise limit}

\date{\today}

\author{Jonathan Demaeyer}
\author{Pierre Gaspard}
\affiliation{Center for Nonlinear Phenomena and Complex Systems,\\
Universit\'{e} libre de Bruxelles (U.L.B.), Code Postal 231, Campus Plaine, B-1050 Brussels, Belgium}

\begin{abstract}
The effect of noise is studied in one-dimensional maps undergoing transcritical, tangent, and pitchfork bifurcations.  The attractors of the noiseless map become metastable states in the presence of noise.
In the weak-noise limit, a symplectic two-dimensional map is associated with the original one-dimensional map.  The consequences of their noninvertibility on the phase-space structures are discussed.  Heteroclinic orbits are identified which play a key role in the determination of the escape rates from the metastable states. Near bifurcations, the critical slowing down justifies the use of
a continuous-time approximation replacing maps by flows, which allows the analytic calculation of the escape rates.  This method provides the universal scaling behavior of the escape rates at the bifurcations.
\end{abstract}


\maketitle

\section{Introduction \label{intro}}

At the macroscale, deterministic dynamics often emerges out 
of the random motion of the many subunits interacting in a complex system. 
In between the micro- and macroscales, the time evolution may admit
a mesoscopic description in terms of stochastic processes in which the deterministic dynamics is affected by noise.  In such systems, the deterministic dynamics rules the mean fields or slow modes emerging out of the many rapidly fluctuating degrees of freedom.  At the mesoscopic scale, these latter
can still manifest themselves and perturb the deterministic time evolution of the slow modes by fast
additive or multiplicative noises \cite{vK81,G04}.  Such stochastic processes have been considered for Brownian motion \cite{L08}, mesoscopic electric circuits \cite{N28} and superconducting junctions \cite{DEMCC87}, mesoscopic lasers and other optoelectronic devices \cite{G91}, fluctuating hydrodynamics \cite{LL90,OS06}, nucleation processes \cite{TO98}, reaction rate theory \cite{HTB90}, nonequilibrium chemical kinetics \cite{NP77,G02}, as well as collective behaviors such as synchronization, flocking or swarming in biology \cite{K84,PRK01,GC04,LGNS04,CDFSTB01,DAGPD86,YEECBKMS09}.

The presence of noise may modify qualitatively the time evolution of systems.  In particular, noise can activate the crossing of otherwise impenetrable barriers.  Therefore, states which are stable with
respect to the deterministic dynamics may become metastable under the effect of noise, as it is the case in nucleation processes or chemical reactions \cite{TO98,HTB90,NP77}.  Instead of remaining forever in a stable state, the trajectory thus escapes from the metastable state with a rate, which corresponds to the lifetime of the system in the metastable state. The phenomenon of noise-induced escape is known in many systems including chemical reactions \cite{HTB90}, nucleation processes \cite{TO98}, Josephson tunnel junctions \cite{DEMCC87}, semiconductor lasers \cite{HZRD00}, or systems with chaotic attractors or fractal basin boundaries \cite{B89,KG04,KG04bis,ES08,BMLSMcC05,SBLMcC05}.

Analytical methods have been developed to evaluate the lifetime of metastable states, especially, in the weak-noise limit.  In this limit, the lifetime is often observed to grow exponentially with the inverse of the noise amplitude, a behavior which is reminiscent of Arrhenius' law in chemical kinetics.  In microscopic chemical reactions, the noise is of thermal origin and its amplitude is controlled by the temperature so that the rate of hopping over the energy barrier is proportional to a Boltzmann factor at the ambient temperature. In macroscopic complex systems, the fluctuations come from the individual motions of the multiple active entities interacting in the system so that the noise amplitude is related to the size of the population of entities and has no thermal origin.  In such complex systems, the analogue of the activation energy is given by the action functional ruling the stochastic process in the weak-noise limit.  

If the noise is weak enough, a variational principle can be used to deduce a Lagrangian or Hamiltonian dynamical system from the action functional, as shown by Onsager and Machlup \cite{OM53}, Freidlin and Wentzell \cite{FW84}, and others \cite{MS96,D99,CSPVD99,PVV01,G02JSP}.  Remarkably, this dynamical system is deterministic and symplectic, i.e., area-preserving in problems with one degree of freedom.  In this respect, the method based on this Hamiltonian dynamics is called the symplectic approach.  We notice that this approach is similar to the semiclassical method of quantum mechanics in the limit where the mechanical action is much larger than Planck's constant.  As in quantum mechanics, the action functional plays a key role in the evaluation of various properties of interest \cite{P88,G90,R03}.  

In the present paper, our purpose is to use the symplectic approach to calculate analytically the rate of noise-induced escape from attractors undergoing transcritical or pitchfork bifurcations \cite{N95}.
Such bifurcations occur in dissipative deterministic systems controlled by tuning parameters.

At a transcritical bifurcation, two steady state cross each other and exchange their stability.  
On both sides of the bifurcation, the unstable state may form the boundary with respect to a remote attractor. In this regard, this is a {\it boundary bifurcation}.  In the presence of noise, this boundary can be crossed and trajectories may escape towards the remote attractor. Near the bifurcation, the proximity of the boundary to the stable state can strongly enhance the escape rate in a way to be determined.

At a pitchfork bifurcation, a stable state becomes unstable, leading to the emergence of a pair of stable states.  Therefore, the deterministic system becomes bistable with two attractors beyond the bifurcation.  Noise will induce random across the unstable state forming the boundary between both attractors.
Here also, the escape rate from each one of both attractors is deeply modified close to the bifurcation.

As we show in the present paper, the symplectic approach allows us to evaluate the effects of these bifurcations on the lifetimes of the metastable states in the weak-noise limit.  To investigate these effects, we consider one-dimensional maps with additive independent Gaussian random variables at each time step.  In the weak-noise limit, we obtain a symplectic area-preserving map associated with the action functional of this Gaussian stochastic process.  Near the bifurcation, we show that an appropriate rescaling of time transforms the two-dimensional symplectic map into a continuous-time Hamiltonian system suitable for analytical calculations.  In this way, the noise-induced escape rate can be evaluated analytically.

The paper is organized as follow. In Section \ref{nmap}, we briefly present the escape problem for a general one-dimensional noisy map. In Section \ref{pint}, we detail the path integral formalism for such noisy systems. Thereafter, we perform semiclassical calculation using the steepest-descent method in order to obtain the general expression of the aforementioned symplectic map. Its phase portraits are analyzed for a specific example. With the help of this analysis, we introduce an analytical model in Section \ref{amod} which allows us to calculate an approximate but analytic expression for the analogue of the activation energy in our context. In Section \ref{res}, we present the results given by this model for several noisy maps and compare them with the results of Monte-Carlo simulations. Finally, we give our conclusions in the last section.

\section{Noisy maps \label{nmap}}

\begin{figure}
\includegraphics[width=0.45\textwidth]{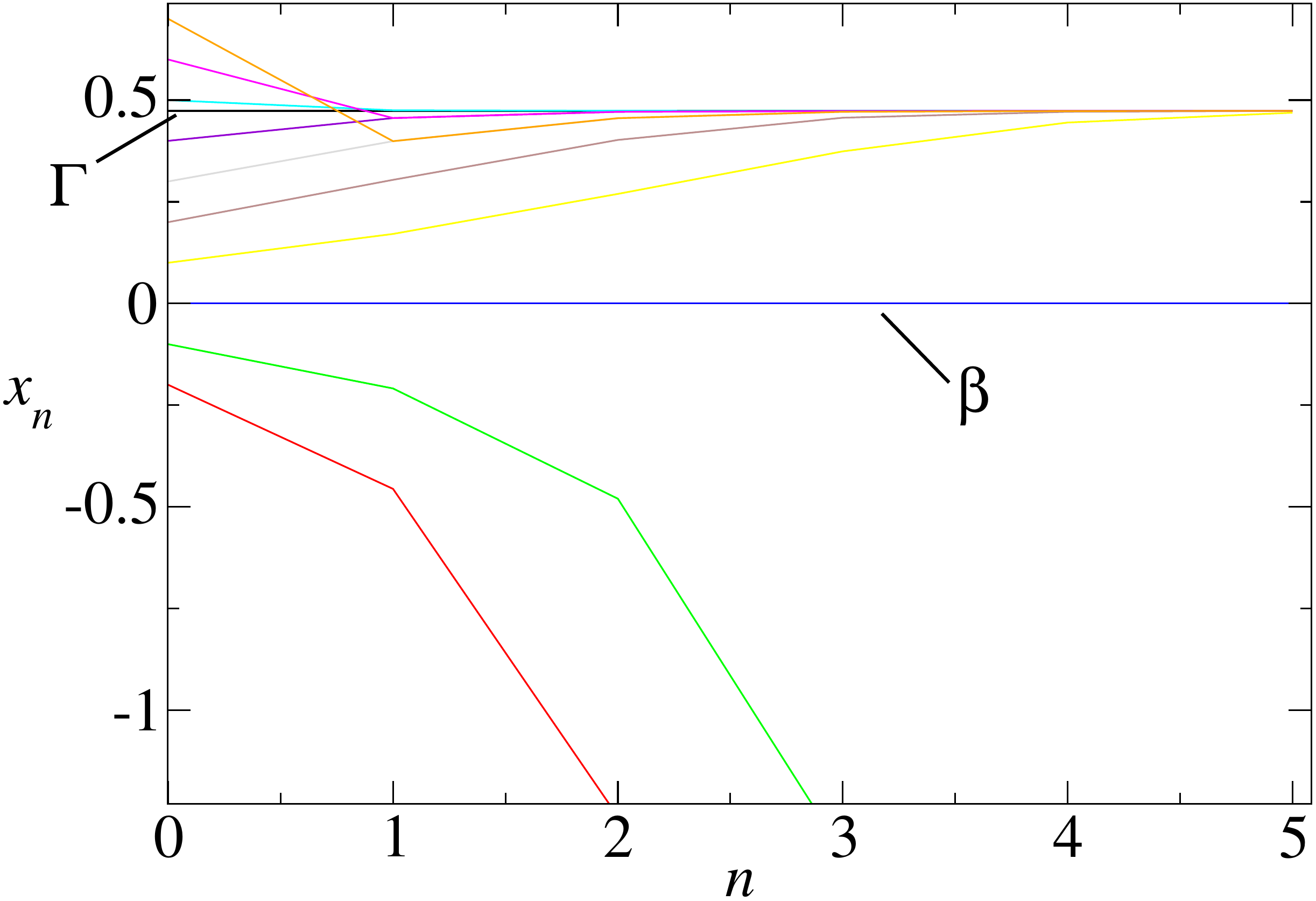}
\caption{\label{dtraj} Phase space $\mathbb{R}$ versus time with typical trajectories of the logistic map $x_{n+1} = \mu x_n (1 - x_n)$ for the parameter value $\mu=1.9$. The attracting fixed point $\Gamma$ of coordinate $x_\Gamma = 1-1/\mu$ and the repulsive fixed point $\beta$ ($x_\beta = 0$) are plotted as straight lines. $\beta$ is one of the two boundaries of the basin of attraction $B(\Gamma) = [0,1]$ of the attractor $\Gamma$. The trajectories outside of $[0,1]$ tend to $-\infty$. Initial conditions $x_0>1$ are sent to the region $]-\infty,0[$ at the first step. Thereafter, these trajectories stay there and tend to $-\infty$.}
\end{figure}

Consider a general one-dimensional map
\begin{equation}\label{eqmap}
x_{n+1} = f(x_n)
\end{equation}
from the set $\mathbb{R}$ of real numbers onto itself.  Such maps are models of the Poincar\'e recurrence in strongly dissipative systems. We notice that the dynamics is deterministic in the sense that each initial condition generates a unique trajectory: $x_n=f^n(x_0)$.  Typically, the trajectories of such deterministic dynamical systems tend to an attractor or escape to infinity (see Fig. \ref{dtraj}).
The attractors of one-dimensional maps can be stationary, periodic, or chaotic.
The basin of attraction $B(\Gamma) \subset \mathbb{R}$ of an attractor $\Gamma$ is the set of all the initial conditions of trajectories converging to $\Gamma$. 

\begin{figure}
\includegraphics[width=0.45\textwidth]{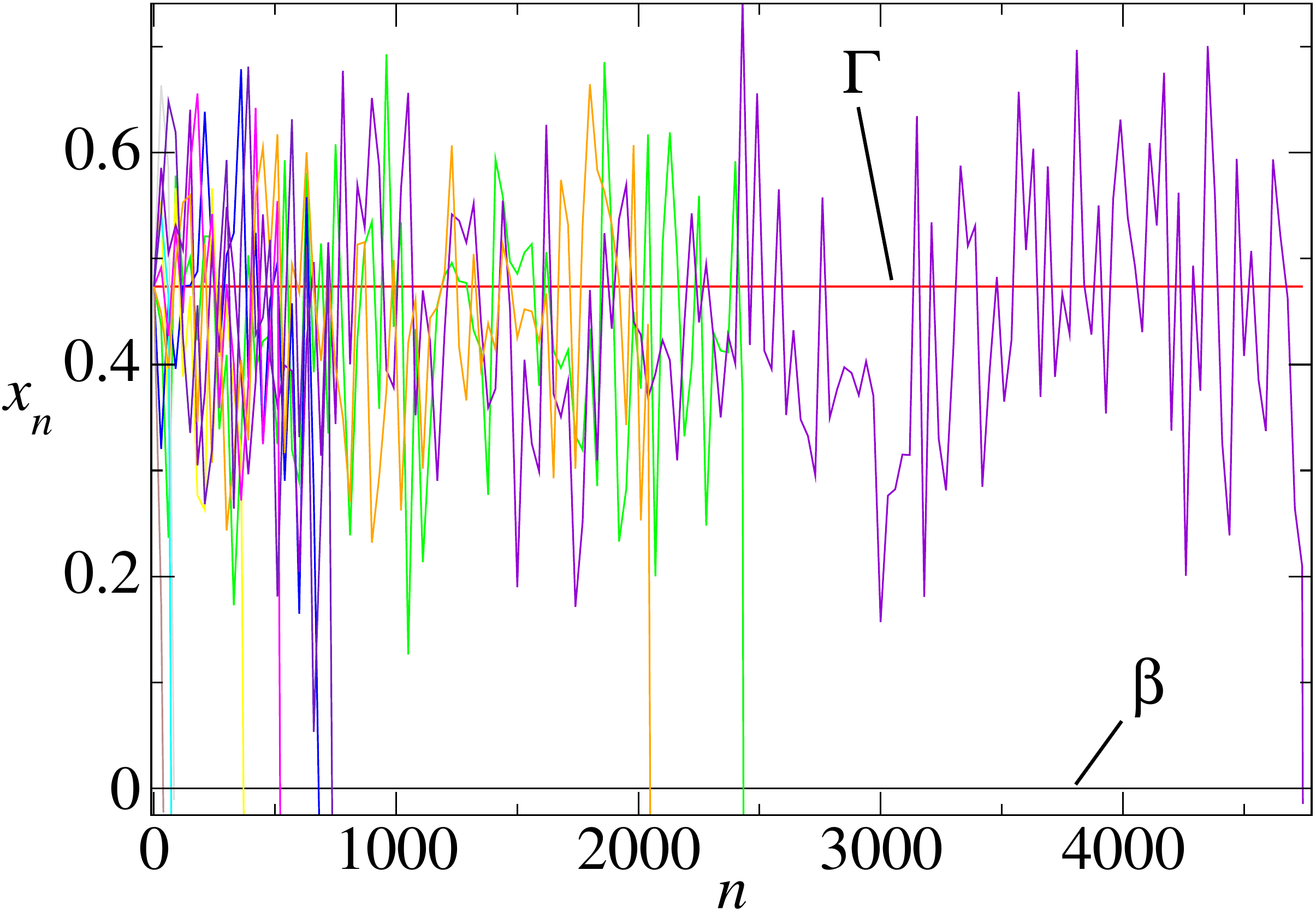}
\caption{\label{ntraj} Same plot as Fig. \ref{dtraj} but for the noisy logistic map (\ref{eqstoch}) with $\mu=1.9$.  For the sake of clarity, each trajectory is plotted each thirty steps. Each trajectory starts on the attracting fixed point $\Gamma$ and fluctuates for a while before crossing the boundary $\beta$ of the basin of attraction $B(\Gamma) = [0,1]$. After this crossing, the trajectory is driven to $-\infty$.}
\end{figure}

In some circumstances, the deterministic dynamics is affected by additive noise
and the quantity $x$ is ruled by the following noisy map
\begin{equation}\label{eqstoch}
x_{n+1} = f(x_n) + \xi_n
\end{equation}
where $\{\xi_n\}_{n=-\infty}^{+\infty}$ is a sequence of independent Gaussian random variables
\begin{eqnarray}
\langle \xi_n \rangle & = & 0 \label{aver_G}\\
\langle \xi_n \xi_m \rangle & = & \varepsilon \, \delta_{nm} \label{corr_G}
\end{eqnarray}
$\varepsilon$ being the amplitude of the noise and $\delta_{nm}$ the Kronecker symbol.
The trajectories of this system are no longer deterministic since the trajectories $\{x_i\}_{i=0}^n$ depends on the sequence of random variables $\{\xi_i\}_{i=0}^{n}$ besides the initial condition $x_0$.

One of the consequences is that the noise perturbs the stability of the attractors.  The boundaries of the basin of attraction $B(\Gamma)$ can be crossed and the trajectory may escape from attractors. Therefore, the attractors $\Gamma$ of the deterministic dynamics typically become metastable in the presence of noise.  Noise-induced escape is illustrated in Fig. \ref{ntraj} for a noisy map with an otherwise stable steady state.

\subsection{Perron-Frobenius equation and operator}

Although the trajectories of noisy maps are random,
a deterministic description is recovered for the probability density $\rho_n(x)$
that the system is found in the position $x$ at the $n^{\rm th}$ time step.
Indeed, the probability density is ruled by the Perron-Frobenius equation~\footnote{When none are specified, the domain of integration is the whole state space. For example, here it is $\mathbb{R}$.} given by
\begin{equation}\label{PFnoise}
\rho_{n+1} (y) = \int K(x,y) \, \rho_n (x) \, \mathrm{d}x
\end{equation}
in terms of a stochastic kernel $K(x,y)=g[y-f(x)]$ where $g(\xi)$ is the probability density of the noise \cite{LM85}. For the Gaussian noise (\ref{aver_G})-(\ref{corr_G}) of noise amplitude $\varepsilon$,
this probability density is given by
\begin{equation}\label{stokerwn}
g(\xi)=\frac{1}{\sqrt{2 \pi \varepsilon}} \, \exp\left(-\frac{\xi^2}{2 \varepsilon}\right)
\end{equation}

In the zero-noise limit $\varepsilon\to 0$, the Gaussian probability density tends to the Dirac delta distribution and the Perron-Frobenius equation of the deterministic map is recovered \cite{LM85}:
\begin{equation}
\rho_{n+1} (y) = \int \delta\left[y-f(x)\right] \rho_n (x) \, \mathrm{d}x
\end{equation}

We notice that, with or without noise, the Perron-Frobenius equation is defined by a linear operator $\hat P$, which preserves the non-negativity of the probability density as well as its normalization to unity: $\int \rho_n(x) \, \mathrm{d}x=1$. In order to analyze the long-time behavior of the probability density, an eigenvalue problem can be posed for the Perron-Frobenius operator 
\begin{equation}
\hat P \Psi_{\alpha}(x) = \chi_{\alpha}  \Psi_{\alpha}(x)
\end{equation}
Since the probability density is not expected to grow without bound, 
the eigenvalues should belong to the unit disk: $\vert\chi_{\alpha}\vert \leq 1$.

The border of the unit disk contains the single eigenvalue $\chi_0=1$ if the probability density converges to a unique stationary solution of the Perron-Frobenius equation.
In this case, all the other eigenvalues are strictly inside the unit disk, $\vert\chi_{\alpha}\vert < 1$ for $\alpha\neq 0$.  As a consequence, the components of the probability density corresponding to
these other eigenvalues are exponentially damped in the long-time limit, allowing the convergence
of the probability density to the eigenfunction $\Psi_0$ associated with the leading eigenvalue $\chi_0=1$.

In contrast, the probability density is expected to vanish if trajectories escape to infinity
as observed in Fig. \ref{ntraj}.  In this further case, the border of the unit disk does not contain any eigenvalue, otherwise the asymptotic probability density would not be vanishing.
Instead, the long-time behavior is dominated by the eigenvalue which is the closest to the border of the unit disk: $\chi_0 <1$ such that $\vert\chi_{\alpha}\vert < \vert\chi_0\vert$ for $\alpha\neq 0$.
After $n$ iterates of the Perron-Frobenius operator, the probability density is damped by a factor $\chi_0^n$ so that we may expect an exponential decay in this situation.
The corresponding eigenfunction $\Psi_0(x)$ is non-negative according to the Perron-Frobenius theorem.

\subsection{Escape rate and mean escape time}
\label{subsec:esc_rate}

Thanks to the probability density $\rho_n(x)$, we can write down the probability that the escape from the attractor $\Gamma$ occurs at a time greater than $n$ in the noisy system (\ref{eqstoch}):
\begin{equation}\label{surv_proba}
P(\{t>n\}) \equiv G_n = \int_{B(\Gamma)} \rho_n (x) \, \mathrm{d}x
\end{equation}
where $B(\Gamma)$ is the basin of attraction of $\Gamma$.
This probability is also called the survival probability.
On the other hand, the probability to escape at a time $n$ is given by
\begin{equation}\label{tescdist}
P(\{t=n\}) \equiv P_n = G_{n-1} - G_n
\end{equation}
We notice that the normalization condition $\sum_{n=1}^{\infty} P_n =G_0=1$ is satisfied.

If trajectories escape to infinity, the escape rate $\gamma$ 
of the attractor of $\Gamma$ is defined as the long-time limit
of the exponential decay rate of the survival probability:
\begin{equation}\label{esc_rate}
\gamma \equiv \lim_{n\to\infty} -\frac{1}{n}\ln G_n
\end{equation}
In this case, the escape can alternatively be defined in terms of the leading eigenvalue $\chi_0$ of the Perron-Frobenius operator.  Since the probability density is damped as $\chi_0^n$ for $n\to\infty$, so does the survival probability (\ref{surv_proba}) and we infer that
\begin{equation}\label{esc_rate_chi_0}
\gamma = - \ln \chi_0
\end{equation}
The inverse of the escape rate is the lifetime of the metastable state corresponding to the eigenfunction $\Psi_0(x)$: 
\begin{equation}
\tau \equiv \frac{1}{\gamma}
\end{equation}

On the other hand, the {\it mean escape time} can be defined as the statistical average of the random time the trajectory escapes from the basin $B(\Gamma)$:
\begin{equation}\label{esc_time}
\langle t \rangle \equiv \sum_{n=1}^{\infty} n \, P_n = \sum_{n=0}^{\infty} G_n
\end{equation}
If the decay is exponential in the long-time limit, $G_n \simeq {\rm e}^{-\gamma n}$,
the mean escape time is given by $\langle t\rangle \simeq (1-{\rm e}^{-\gamma})^{-1}$.
In the weak-noise limit where the escape rate is expected to vanish as
\begin{equation}\label{teedep}
\gamma \sim \exp \left(-\frac{W_0}{\varepsilon}\right)
\end{equation}
for $\varepsilon\to 0$, the mean escape time can be related to the escape rate itself by the simple relation
\begin{equation}\label{rate_time}
\lim_{\gamma\to 0} \gamma \, \langle t \rangle = 1
\end{equation}
Otherwise, the mean escape time is not simply given as the inverse of the escape rate because, in general, the mean escape time depends on the transients before the long-time exponential decay at the escape rate. 

\subsection{Relaxation rate and mean first-exit time}
\label{first-exit_time}

If trajectories do not escape to infinity but randomly jump between two attractors of the deterministic dynamics, the probability density is expected to converge towards a stationary density given by the eigenfunction $\Psi_0(x)$ of the Perron-Frobenius operator corresponding to the eigenvalue $\chi_0=1$.
In this case, the escape rate (\ref{esc_rate}) vanishes and the quantity of interest is given by the next-to-leading eigenvalue $\chi_1$ such that $\vert\chi_{\alpha}\vert < \chi_1 < 1$ for $\alpha\neq 0,1$ if this latter is unique.  The eigenvalue $\chi_1$ controls the convergence of the probability density towards the stationary state.  This convergence is exponential, which defines the relaxation rate
\begin{equation}\label{relax_rate}
\tilde\gamma = - \ln \chi_1
\end{equation}
This relaxation rate can be interpreted as the rate of jumps between the two attractors due to the noise.
In some cases, this relaxation rate can be defined in the presence of escape if there is a clear separation of time scales between both.

A mean first-exit time can also be defined as the statistical average of the random time of first exit from a given basin of attraction $B(\Gamma)$.  In this scheme, the trajectories starts with an initial condition inside the basin $B(\Gamma)$, evolves in time according to the noisy map (\ref{eqstoch}), and stops as soon as the trajectory escapes from the basin $B(\Gamma)$.  The Perron-Frobenius operator $\hat Q$ of this first-exit process is defined by restricting the domain of integration to the basin $B(\Gamma)$:
\begin{equation}\label{PFnoise_Q}
\rho_{n+1} (y) = \int_{B(\Gamma)} K(x,y) \, \rho_n (x) \, \mathrm{d}x
\end{equation}
For a bounded basin $B(\Gamma)$, all the eigenvalues of this further operator are strictly inside the unit disk and an escape rate $\gamma$ can be defined by Eq. (\ref{esc_rate_chi_0}) in terms of the leading eigenvalue $\chi_0$, as in the previous Subsection \ref{subsec:esc_rate}.  Furthermore, a mean escape time $\langle t\rangle$ can be defined for the present process by Eq. (\ref{esc_time}).
If the attractor $\Gamma$ of the deterministic dynamics is strictly inside its basin of attraction $B(\Gamma)$, we expect that, in the weak-noise limit, the relaxation rate should behave as
$\tilde\gamma \sim \exp(-W_0/\varepsilon)$, for $\varepsilon\to 0$, and be related to the mean escape time $\langle t \rangle$ and the corresponding escape rate $\gamma$ according to
\begin{equation}
\lim_{\tilde\gamma\to 0} \tilde\gamma \, \langle t \rangle = \lim_{\gamma\to 0} \gamma \, \langle t \rangle = 1
\end{equation}

For both escape and relaxation, the determination of the lifetime of the metastable state thus goes by the evaluation of the quantity $W_0$ which is the analogue of the activation energy in systems with thermal noise.

\section{Path integrals and the symplectic approach \label{pint}}

The aim of the present section is to obtain the symplectic map ruling the process in the weak-noise limit.

\subsection{Path integrals}

Starting with an initial density $\rho_0$ and iterating Eq. (\ref{PFnoise}), one can obtain the probability density at time $n$ in the form
\begin{equation}\label{nevol}
\rho_n (x_n) = \int \mathrm{d} x_0 \, K_n (x_0,x_n) \, \rho_0 (x_0)
\end{equation}
where $K_n$ is an integral kernel resulting from $n$ iterations. 
This kernel is the propagator associated with the time evolution of the stochastic process.
For the Gaussian noise (\ref{stokerwn}), this kernel is given by the multiple or path integral
\begin{eqnarray}\label{kerbb}
K_n (x_0,x_n) & = & \frac{1}{(\sqrt{2 \pi \varepsilon})^n} \int \mathrm{d} x_{n-1} \ldots \int \mathrm{d} x_1 \\ \nonumber
 & & \times \exp \left[ - \frac{1}{\varepsilon} \, W_n (x_0,x_1,\ldots,x_{n-1},x_n)\right]
\end{eqnarray}
where
\begin{equation}\label{fact}
W_n (x_0,\ldots,x_n) = \frac{1}{2} \sum_{i=0}^{n-1} \left[x_{i+1} - f(x_i)\right]^2
\end{equation}
which defines the action functional of the Gaussian stochastic process.

\subsection{Symplectic map \label{semiapp}}

In the limit of weak noise $\varepsilon \to 0$, the propagator (\ref{kerbb}) can be integrated by
the steepest-descent method, which will select the extrema of the action functional (\ref{fact}).
These extrema are obtained by applying Hamilton's variational principle for the
action function (\ref{fact}).  The selected paths are thus the trajectories 
of a symplectic deterministic system \cite{FJ05} .
With this Hamiltonian formalism in mind, the functional (\ref{fact}) can be written in terms of a generating function $F$ of the first kind \cite{G50}
\begin{equation}\label{Wpos}
W_n (x_0,\ldots,x_n) = \sum_{i=0}^{n-1} F(x_i, x_{i+1})
\end{equation}
where
\begin{equation}\label{gfunc}
F(x_i,x_{i+1}) = \frac{1}{2} \, \left[x_{i+1} - f(x_i) \right]^2
\end{equation}
Introducing the momentum $p_i$ canonically conjugated to the position $x_i$,
the evolution over one time step is conceived as a canonical transformation corresponding to this generating function
\begin{equation}\label{can_transf}
\left\{ \begin{array}{lcl}
p_{i+1} & = & \frac{\partial F(x_i,x_{i+1})}{\partial x_{i+1}} \\ \\
p_{i} & = & - \frac{\partial F(x_i,x_{i+1})}{\partial x_i}
\end{array} \right.
\end{equation}
With the generating function (\ref{gfunc}), we obtain the two-dimensional map
\begin{equation}\label{mapsymp}
\phi \; \left\{ \begin{array}{lcl}
x_{i+1} & = & f (x_i) + \frac{p_i}{f'(x_i)} \\ \\
p_{i+1} & = & \frac{p_i}{f'(x_i)}
\end{array} \right.
\end{equation}
This map is symplectic because it preserves the symplectic differential two-form 
\begin{equation}
\mathrm{d}x_{i+1} \wedge \mathrm{d}p_{i+1} = \mathrm{d}x_i \wedge \mathrm{d}p_i
\end{equation}
where $\wedge$ denotes the exterior or wedge product \cite{M92}.  Accordingly, the Jacobian of the map takes the unit value, $\det D\phi=1$, so that the map is area-preserving in the two-dimensional phase space $\mathcal{M}=\{(x,p): x\in\mathbb{R},p\in\mathbb{R}\}$.

Such area-preserving maps are expected to be typically chaotic as it is the case for the standard map \cite{M92}.  The multitude of possible trajectories reflects the complexity of the landscape of the multidimensional functional (\ref{Wpos}). This landscape unveils many local extrema among which the local minima should be selected in order to guarantee the local normalization of the probability distribution underlying the propagator (\ref{kerbb}).

We emphasize that the symplectic map (\ref{mapsymp}) has special properties which are specific to the problem at hand.  

First, the map (\ref{mapsymp}) leaves invariant the subspace $p=0$ of the phase space $\mathcal{M}$.  In this subspace, the two-dimensional map (\ref{mapsymp}) reduces to the noiseless one-dimensional dissipative map $x_{i+1} = f (x_i)$.  In consequence, the attracting fixed points of the original map $f$ are hyperbolic fixed points of the symplectic map (\ref{mapsymp}) that lie in the subspace $p=0$. Note that the repulsive fixed points of the map $f$ are also hyperbolic fixed points of (\ref{mapsymp}) and also lie on the subspace $p=0$.

Secondly, the symplectic map (\ref{mapsymp}) is not uniquely invertible and may admit several inverses. 
This is the consequence of the noninvertible character of the one-dimensional map $f$ itself, which often has extrema where its derivative $f'(x)$ vanishes.  Indeed, the inverses of the two-dimensional map (\ref{mapsymp}) are directly associated with the inverses $f_j^{-1}$ of the one-dimensional map $f$, such that $f_j^{-1}\left[f(x)\right]=x$ for $j=1,2,...,m$. The inverses of the two-dimensional map (\ref{mapsymp}) can thus be written as
\begin{equation}\label{invmapsymp}
\phi_j^{-1} \; \left\{ \begin{array}{lcl}
x_i& = & f_j^{-1}(x_{i+1} - p_{i+1}) \\ \\
p_i & = & p_{i+1} \, f'\left[ f_j^{-1}(x_{i+1} - p_{i+1}) \right]
\end{array} \right.
\end{equation}
with $j=1,2,...,m$.  The inverses of one-dimensional map become singular at the positions such that $f'(x_k)=0$ with $k=1,2,...,\tilde m$.  These loci correspond to the borders of the domains where the inverses (\ref{invmapsymp}) are defined:
\begin{equation}\label{border}
x-p=f(x_k)   \qquad \mbox{such that} \qquad f'(x_k)=0
\end{equation}
with $k=1,2,...,\tilde m$.
For noninvertible two-dimensional maps such as Eq. (\ref{mapsymp}) with a vanishing denominator, the concepts of focal point and prefocal curves (or lines) have been introduced \cite{BGM99,BGM03}. They will be defined here below in the discussion of the phase-space structures generated by these maps, which are different in this regard from those of the invertible area-preserving maps such as the standard map \cite{M92}.

\subsection{Phase space\label{phasespaces}}

This section is devoted to the phase-space structures generated by two-dimensional symplectic maps (\ref{mapsymp}) associated with one-dimensional maps $f$ undergoing bifurcations. 
The study of typical phase portraits of these symplectic maps is important because the contributions to the propagator (\ref{kerbb}) are evaluated in the weak-noise limit in terms of the trajectories of these maps. Moreover, statistical averages are related to the propagator and, as we shall see, the study of phase space gives an insight on their properties.

\subsubsection{Invariant subspace $p=0$}

As mentioned here above, the straight line $p=0$ is an invariant subspace of the phase space $\mathcal{M}=\{(x,p): x\in\mathbb{R},p\in\mathbb{R}\}$.  This line is the $x$-axis where the dynamics is ruled by the one-dimensional map $x_{i+1}=f(x_i)$ of the noiseless macroscopic system.
The trajectories outside the invariant subspace are specifically associated with noise.
This invariant subspace divides the phase space in two half planes and, therefore, constitutes a major organizing geometric structure.  

\subsubsection{Fixed points}

The symplectic map (\ref{mapsymp}) has two types of fixed points.

The fixed points $x=f(x)$ of the one-dimensional map, which are found in the invariant subspace $p=0$.
For these fixed points, the linearized map has the eigenvalues $\Lambda_+=f'(x)$ and $\Lambda_-=f'(x)^{-1}$.  Accordingly, these fixed points are hyperbolic (or possibly parabolic in the cases of marginal stability).  If the point is attracting (resp. repelling) for the one-dimensional map, the unstable (resp. stable) direction is transverse to the invariant subspace $p=0$.

On the other hand, the symplectic map may admit genuine fixed points outside the invariant subspace $p=0$ for
\begin{equation}\label{fixed_pts}
\left\{ \begin{array}{l}
p =  x-f(x) \\ \\
f'(x) =  1
\end{array} \right.
\end{equation}
These fixed points have eigenvalues given by the roots of the characteristic equation:
\begin{equation}
\Lambda^2 + \left[ p \, f"(x) - 2\right] \Lambda + 1 = 0
\end{equation}
so that they can be either elliptic or hyperbolic (or also parabolic in the marginal cases).

\subsubsection{Singular lines, focal points, and prefocal lines}

Because of its vanishing denominators,  the two-dimensional map (\ref{mapsymp}) is not defined on the {\it singular lines} 
\begin{equation}
\delta_{{\rm s},k} = \{ (x_k,p): f'(x_k)=0, p\in\mathbb{R}\}
\end{equation}
with $k=1,2,...,\tilde m$.  The implications of these singular lines on the phase-space structures generated by the map have been discovered by Mira and coworkers who introduced the concepts of focal points and prefocal lines \cite{BGM99,BGM03}.

As long as the numerator corresponding to the denominator is not vanishing, the points of the singular lines are sent to infinity by the map (\ref{mapsymp}).  However, this might not be the case if the numerator vanishes with the denominator at the so-called {\it focal points} \cite{BGM99,BGM03}
\begin{equation}
\boldsymbol{z}_k = (x_k,p=0)
\end{equation}

Since the map (\ref{mapsymp}) displays an undetermined ratio $0/0$ at each focal point, the image of the focal point is not necessarily sent at infinity and we may wonder what is the phase-space structure corresponding to this image.  To answer this question, let us consider a line $\gamma$ going through the focal point and defined by the following parametric equations:
\begin{equation}\label{gamma_line}
\gamma \; \left\{ \begin{array}{lcl}
x & = & x_k + \xi  \\ \\
p & = & \alpha \, \xi + O(\xi^2)
\end{array} \right.
\end{equation}
where $\xi$ is the parameter and $\alpha$ the slope of the line at the focal point.
The image of this line is given by
\begin{equation}\label{image_gamma_line}
\phi(\gamma) \; \left\{ \begin{array}{lcl}
x' & = & f(x_k + \xi) + \frac{\alpha \, \xi + O(\xi^2)}{f'(x_k + \xi)} = f(x_k) + \frac{\alpha}{f"(x_k)} + O(\xi) \\ \\
p' & = & \frac{\alpha \, \xi + O(\xi^2)}{f'(x_k + \xi)} = \frac{\alpha}{f"(x_k)} + O(\xi) 
\end{array} \right.
\end{equation}
where we have expanded in powers of the parameter $\xi$ around $\xi=0$, using $f'(x_k)=0$, and assuming that $f"(x_k)\neq 0$.  The image $\phi(\gamma)$ is a line for varying values of the parameter $\xi$.  Taking the limit $\xi=0$, we obtain the image of the focal point:
\begin{equation}\label{image_focal_point}
\phi(\boldsymbol{z}_k) \; \left\{ \begin{array}{lcl}
x' & = & f(x_k) + \frac{\alpha}{f"(x_k)}  \\ \\
p' & = & \frac{\alpha}{f"(x_k)} 
\end{array} \right.
\end{equation}
We notice that the image depends on the slope $\alpha$ given to the line $\gamma$ in Eq. (\ref{gamma_line}), whereupon the images are multiple and form a line called the {\it prefocal line}:
\begin{equation}\label{prefocal}
\delta_{\boldsymbol{z}_k}: \qquad x' = p' + f(x_k) 
\end{equation}
which coincides with the border (\ref{border}) of the domains of definition of the inverse two-dimensional maps (\ref{invmapsymp}).  Accordingly, the prefocal line  is mapped onto the focal point under the inverse map (\ref{invmapsymp}) having the prefocal line in its domain of definition:
\begin{equation}\label{preimage_prefocal}
\phi_j^{-1} \left( \delta_{\boldsymbol{z}_k}\right) = \boldsymbol{z}_k
\end{equation}
In this sense, the prefocal line constitutes the image of the focal point, $\delta_{\boldsymbol{z}_k} = \phi(\boldsymbol{z}_k)$, showing that a point can be mapped onto a line in such two-dimensional map (\ref{mapsymp}) with a vanishing denominator \cite{BGM99,BGM03}.

\subsubsection{Global stable set and unstable manifold}

The structures generated by the map $\phi$ in its phase space $\mathcal{M}$ can be analyzed in terms of the stable and unstable sets associated with the saddle points.  

The unstable set ${\cal W}^{\rm u}(\boldsymbol{z})$ of the hyperbolic fixed point $\boldsymbol{z}=(x,p)$ consists in the set of points that converge to $\boldsymbol{z}$ under backward iterations of $\phi$. This can be expressed in terms of the union of the successive images of the local unstable manifold ${\cal W}_\mathrm{loc}^{\rm u}(\boldsymbol{z})$, i. e., the unstable set in a neighborhood of $\boldsymbol{z}$:
\begin{equation}
{\cal W}^{\rm u}(\boldsymbol{z}) = \bigcup_{n=1}^\infty \phi^n [{\cal W}_\mathrm{loc}^{\rm u}(\boldsymbol{z})]
\end{equation}
Even if $\phi$ is not invertible, the images of the local manifold ${\cal W}_\mathrm{loc}^{\rm u}(\boldsymbol{z})$ will be uniquely determined.  Accordingly, the set ${\cal W}^{\rm u}(\boldsymbol{z})$ is a manifold in the phase space and is called the \emph{unstable manifold} \cite{EKO04}.

On the other hand, the stable set of the hyperbolic fixed point $\boldsymbol{z}$ is defined as the set of points that converge to $\boldsymbol{z}$ under forward iterations of $\phi$. It can be obtained as the union of successive pre-images given by the possibly multiple \emph{inverse maps} $\phi^{-1}_j$ acting on the local stable manifold ${\cal W}_\mathrm{loc}^{\rm s}(\boldsymbol{z})$ :
\begin{equation}
{\cal W}^{\rm s}(\boldsymbol{z}) = \bigcup_{n=1}^\infty \bigcup_j \phi^{-n}_j [{\cal W}_\mathrm{loc}^{\rm s}(\boldsymbol{z})]
\end{equation}
If $\phi$ is invertible, this set is a manifold and one speaks about \emph{the stable manifold} of $\boldsymbol{z}$. However, if there are multiple inverse maps, the global stable set may consist of disjoint pieces and is not a manifold \cite{EKO04,BCGM96}. Since we here consider noninvertible maps $f$, the symplectic map (\ref{mapsymp}) is noninvertible so that we speak about the \emph{global stable set} (instead of \emph{stable manifold}).

\begin{figure}
\includegraphics[width=0.5\textwidth]{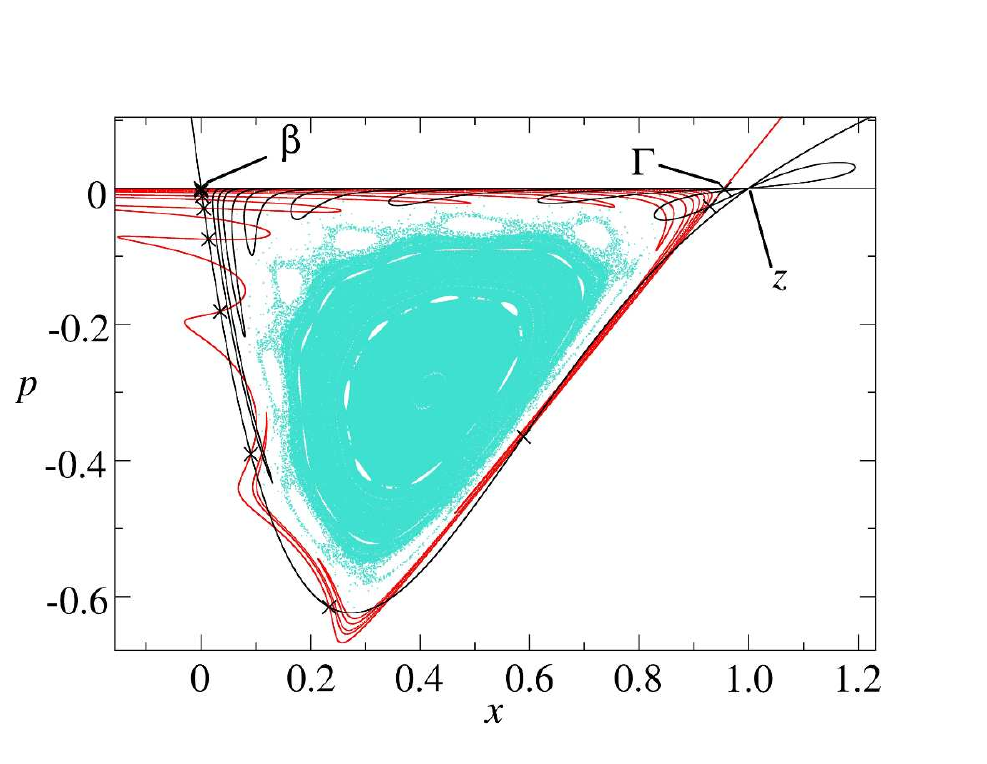}
\caption{\label{expps26} The phase space $\mathcal{M}$ of the symplectic map (\ref{mapsymp}) associated with the exponential map $x_{n+1}=\mu x_n\exp(-x_n)$ with $\mu=2.6$. Global unstable set (in red) and global stable manifold (in black) of respectively the fixed point $\Gamma$ and the boundary $\beta$. At the intersections of the two sets, the heteroclinic trajectory (black cross) between these two fixed points. In light blue, some bounded trajectories inside the region delimited by the two sets.}
\end{figure}

At the intersection of the global stable set and the unstable manifold associated with one and the same hyperbolic fixed point $\boldsymbol{z}$, we find the so-called {\it homoclinic orbit} ${\cal W}^{\rm s}(\boldsymbol{z})\cap {\cal W}^{\rm u}(\boldsymbol{z})$.

If the global stable set and the unstable manifold are attached to different fixed points, we are in the presence of a so-called {\it heteroclinic orbit} ${\cal W}^{\rm s}(\boldsymbol{z})\cap {\cal W}^{\rm u}(\boldsymbol{z}')$, which plays an important role in the following.

As an example, a typical phase portrait of the symplectic map (\ref{mapsymp}) is depicted in Fig. \ref{expps26} for the exponential map $x_{n+1} = \mu x_n \exp(-x_n)$. For the noisy exponential map with the selected value of the parameter $\mu$, the escape scenario is the same as the one detailed in Fig. \ref{ntraj} for the logistic map. Starting from the attractive fixed point $x_\Gamma=\ln \mu$, trajectories escape after crossing of the boundary $x_\beta = 0$. Once this boundary is crossed, the trajectory tends to $-\infty$. We notice that the fixed points of the one-dimensional map correspond to the fixed points $\Gamma=(\ln\mu,0)$ and $\beta=(0,0)$ of the two-dimensional map.  Moreover, the stable set ${\cal W}^{\rm s}(\Gamma)$ and the unstable set  ${\cal W}^{\rm u}(\beta)$ are contained in the invariant subspace $p=0$. The construction of the unstable manifold ${\cal W}^{\rm u}(\Gamma)$ and the global stable set ${\cal W}^{\rm s}(\beta)$ seen in Fig. \ref{expps26} have first required the determination of the manifolds directly emanating from the hyperbolic fixed points. This has been performed with an algorithm specially devoted to noninvertible maps and called the search circle algorithm \cite{EKO04,KO98}. Thereafter, we were able to compute significant parts of the global invariant sets using backward iterations with $\phi^{-1}_j$ for the global stable set ${\cal W}^{\rm s}(\beta)$ and forward iterations with $\phi$ for the unstable manifold ${\cal W}^{\rm u}(\Gamma)$.
We have also computed with precision the heteroclinic orbit ${\cal W}^{\rm s}(\beta)\cap{\cal W}^{\rm u}(\Gamma)$ using an algorithm described in Ref. \cite{Y98}.
An attentive examination of Fig. \ref{expps26} shows that the global stable set ${\cal W}^{\rm s}(\beta)$ of the boundary fixed point $\beta$ forms closed loops around the focal point $\boldsymbol{z}=(1,0)$. This feature is inherent to the map (\ref{mapsymp}) which is not defined on the singular lines where the derivative of the one-dimensional map $f$ vanishes. For the exponential map, there is such a singular line at $x=1$ and its intersection with the subspace $p=0$ where there is a $0/0$ limit gives the focal point $\boldsymbol{z}=(1,0)$ where $f'(1)=0$. The closed loops of the global stable set ${\cal W}^{\rm s}(\beta)$ are attached to this focal point. This feature is related to the noninvertibility of $f$ for which the  focal points divide the configuration space $x$ into different pre-image regions.
Since the nondefinition sets are lines that divide the phase space into distinct regions, the only way for a curve that is forward invariant to cross these lines is to pass through focal points, as here observed for the global stable set ${\cal W}^{\rm s}(\beta)$. We refer the reader to the papers \cite{BGM99,BGM03,EKO05} for more information about this issue.  

\subsubsection{Dynamics close to the bifurcation}

\begin{figure*}
\includegraphics[width=0.9\textwidth]{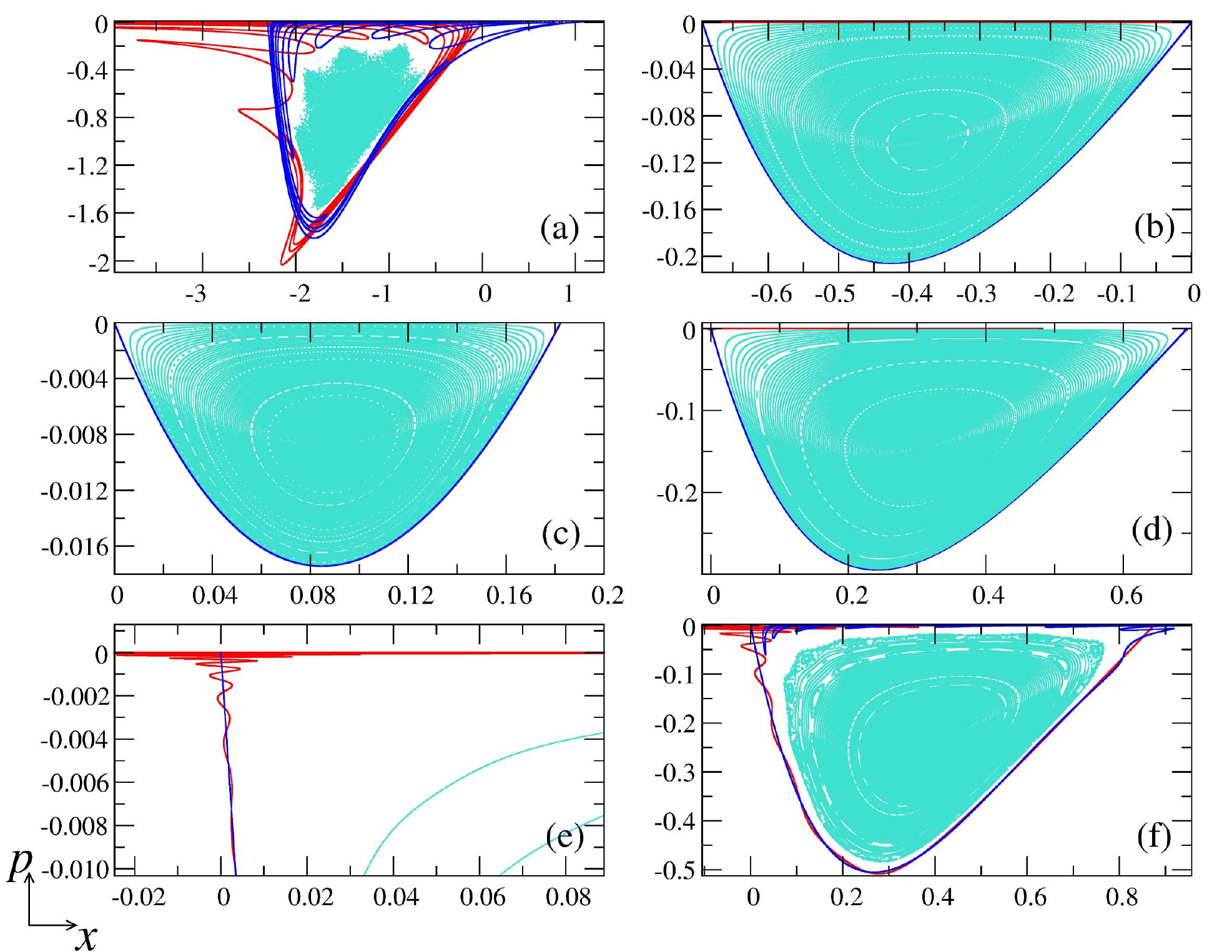}
\caption{\label{exp1pp} Phase portraits of the symplectic map (\ref{mapsymp}) associated with the exponential map $x_{n+1} = \mu x_n \exp(-x_n)$ for different values of the parameter $\mu$ close to the transcritical bifurcation at $\mu=1$: (a) $\mu=0.1$; (b) $\mu=0.5$; (c) $\mu=1.2$; (d) $\mu=2.0$; (f) $\mu=2.4$. The panel (e) is a zoom of (d) around the fixed point $(0,0)$.}
\end{figure*}

Close to the bifurcation, the dynamics undergoes a critical slowing down.
As a consequence, the symplectic map (\ref{mapsymp}) performs small steps at each iteration
in the phase-space domain where the fixed points bifurcate.  It turns out that the dynamics of the map
is similar to a continuous-time flow.  In order to illustrate this phenomenon, we continue to consider
the exponential map $x_{n+1} = \mu x_n \exp(-x_n)$ and we depict in Fig.~\ref{exp1pp} its phase portrait close to the transcritical bifurcation it undergoes at the parameter value $\mu=1$.  At this bifurcation, the two fixed points $x_{\beta}=0$ and $x_{\Gamma}=\ln\mu$ cross each other and exchange their stability.  The unstable of these fixed points is the boundary of the basin of attraction of the other so that a boundary bifurcation happens at the transcritical bifurcation.  Below and above the bifurcation, an elliptic island exists in the lower half plane $p<0$ around the extra fixed point approximately located at $x\simeq (\mu-1)/2$ and $p\simeq-(\mu-1)^2/4$.  At the bifurcation, the three fixed points coalesce at the origin with the surrounding island.

Figures ~\ref{exp1pp}(a) and (f) depicts the phase portraits for values of $\mu$ far from the bifurcation
where we observe that the global stable set and the unstable manifolds are well developed. The trajectories in light blue are trapped and form an elliptic island composed of invariant circles and chaotic zones of moderate extension.  At these values, the system (\ref{mapsymp}) is clearly not close to integrability since the global stable set and the unstable manifold form typical heteroclinic structures.

For values of $\mu$ closer to the bifurcation, Figs.~\ref{exp1pp}(b) and (c) show that most of the trajectories of the elliptic islands are pretty regular lines bounded by the global stable set and the unstable manifold nearly forming a separatrix. The chaotic zones have disappeared on tiny scales in phase space.  Hence, we can conclude that, close to the bifurcation, the system (\ref{mapsymp}) tends to complete integrability.

It is interesting to understand how the transition occurs between the two situations. In Fig.~\ref{exp1pp}(d), the phase space is depicted for an intermediate value between these of Figs.~\ref{exp1pp}(c) and (f). Although the phase portrait is similar to the one in Fig.~\ref{exp1pp}(c), Fig.~\ref{exp1pp}(e) shows a zoom of the global stable  set and the unstable manifold around the origin revealing heteroclinic structures typical of non-integrable dynamics. 

We conclude that, as the bifurcation is approached, the slowing down of the dynamics tends to transform the discrete-time map into a continuous-time flow with one degree of freedom which is thus integrable at the bifurcation. This idea will be elaborated upon in the next section.

\section{Continuous-time limit at bifurcation \label{amod}}

\subsection{Hamiltonian flow}

In this section, we establish the correspondence between the symplectic map (\ref{mapsymp}) and a continuous-time flow with one degree of freedom in the neighborhood of the bifurcation.   This correspondence provides us with a method to compute analytically the properties of the noisy map close to the bifurcation and, in particular, the rate of noise-induced escape from the attractor.

With this aim, we consider a general map of the form:
\begin{equation}\label{fanaexp}
f(x) = \sum_{i=0}^\infty c_i \, x^i
\end{equation}
where the coefficients $c_i(\mu)$ depend on the parameter $\mu$.
For the transcritical and pitchfork bifurcations,  we take $c_0 = 0$ which implies that $x=0$ remains a fixed point across the bifurcation.  For a bifurcation to happen, we must moreover suppose that the map $f$ possesses at least another fixed point. This means that the series (\ref{fanaexp}) contains nonlinear terms. Let us suppose that $c_k$ with $k>1$ is the first non-vanishing coefficient besides $c_1$.

A bifurcation occurs if the fixed point $x=0$ looses or gains stability. This is the case at the value of $\mu$ where the linear stability eigenvalue of the map $f$ at $x=0$ is equal to unity. Since the linear stability of $x=0$ is $f'(0) = c_1$, the bifurcation occurs at the critical parameter value $\mu=\mu_{\rm c}$ where $c_1(\mu_{\rm c})=1$. Near the bifurcation, the dynamics of the map is slowed down since its slope is close to unity, $c_1\simeq 1$, which limits the iterations to small steps.  Consistently,
the other fixed points of the one-dimensional map are approximately given by $x\simeq [(1-c_1)/c_k]^{1/k}$, which coalesce onto the origin $x=0$ at the bifurcation.

In this regard, we can suppose that trajectories $\{x_n\}_{n=-\infty}^{+\infty}$ of the map can be seen as trajectories $x(t)$ of a continuous-time stochastic dynamical system with the time $t=n$.
Since the time step is unity, $\Delta t=1$, differences become derivatives:
\begin{eqnarray}
x_{n+1}-x_n = & = & \left[{\rm e}^{\partial_t} x(t)-x(t)\right]_{t=n} \nonumber \\
 & = & \dot x + \frac{1}{2} \, \ddot x + \, \cdots
\end{eqnarray}
The critical slowing down  near the bifurcation justifies the truncation of this expansion
to the term with the first derivative with respect to the continuous time $t=n$, which amounts to assimilate  the difference between two successive time steps with a time derivative.

To implement this idea, the position $x_n$ at the previous step is subtracted from the noisy map (\ref{eqstoch}), which is rewritten as
\begin{equation}
x_{n+1} - x_n = f(x_n) - x_n + \xi_n
\end{equation}
Introducing the function
\begin{equation}\label{ftrans}
g(x) \equiv f(x) -x 
\end{equation}
the noisy map can be replaced by the Ito stochastic differential equation
\begin{equation}\label{eqstoch2}
\frac{\mathrm{d}x}{\mathrm{d}t} = g (x) + \eta(t)
\end{equation}
with the Gaussian white noise
\begin{eqnarray}
\langle \eta(t) \rangle & = & 0 \\
\langle \eta(t) \eta(t') \rangle & = & \varepsilon \, \delta(t-t')
\end{eqnarray}
We notice that the time integral of this noise defines the Wiener process ${\rm W}(t)=\int_0^t \eta(t')\mathrm{d}t'$.  We emphasize the replacement of the noisy map by the stochastic differential equation is justified by the critical slowing down of the dynamics close to the bifurcation.

\begin{figure*}
\includegraphics[width=\textwidth]{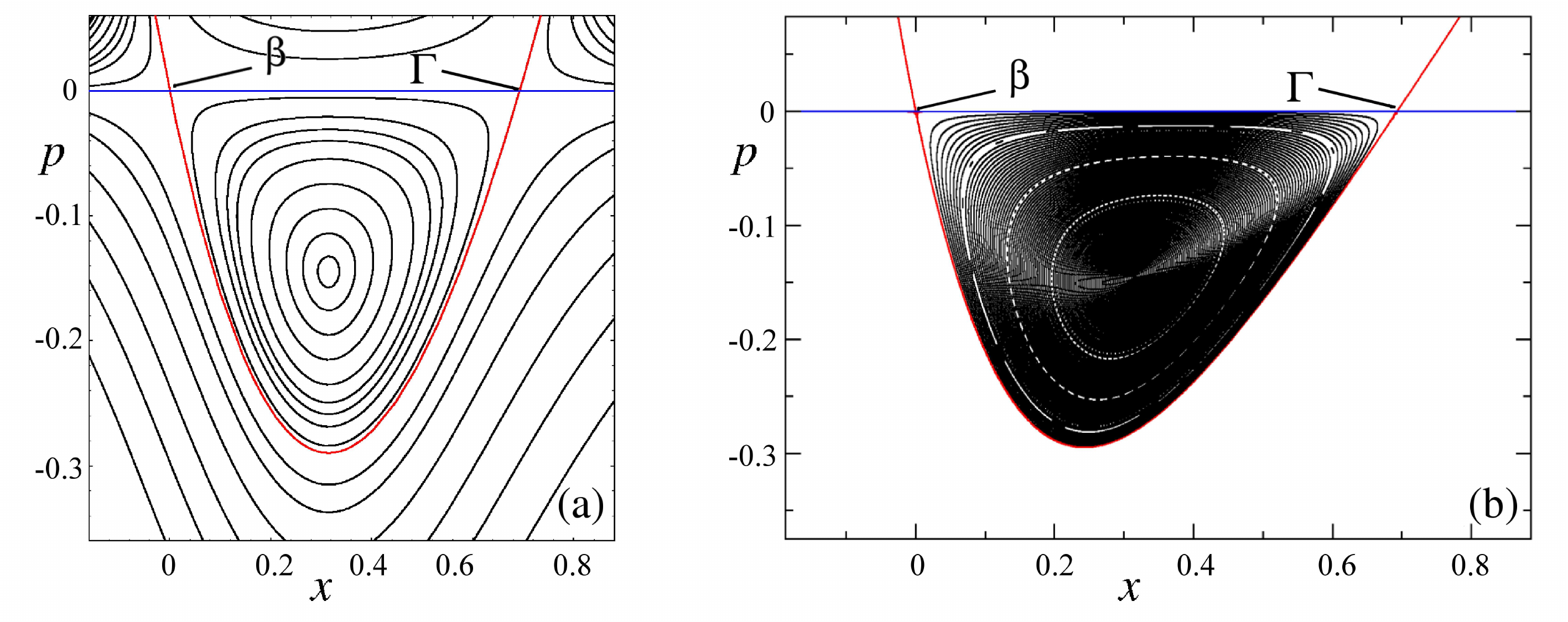}
\caption{\label{ppcdexp1} Comparison between the phase portrait of (a) the Hamiltonian flow (\ref{flowsymp}) and (b) the phase portrait of the symplectic map (\ref{mapsymp}) for the exponential map $x_{n+1} = \mu x_n \exp(-x_n)$ at $\mu = 2$.}
\end{figure*}

In the weak-noise limit, this stochastic process can be analyzed thanks to the theory by Onsager and Machlup \cite{OM53} or Freidlin and Wentzell  \cite{FW84} in terms of a Hamiltonian flow with one degree of freedom.  Its Hamiltonian function is given by
\begin{equation}\label{Hamiltonian}
H= \frac{1}{2} \, p^2 + g(x) \, p 
\end{equation}
This Hamiltonian function rules the Hamilton-Jacobi equation derived from the Fokker-Planck equation associated to the equation (\ref{eqstoch2}).
The trajectories are now solutions of Hamilton's equations
\begin{equation}\label{flowsymp}
\left\{ \begin{array}{lcl}
\dot x & = & g (x) + p \\ \\
\dot p & = & -g' (x) \, p
\end{array} \right.
\end{equation}
where $p$ is the momentum variable canonically conjugated to the position $x$. We notice that, as for the symplectic map (\ref{mapsymp}), the fixed points of the flow $\dot x = g(x)$ are hyperbolic fixed points of Eqs. (\ref{flowsymp}) that lie on the invariant subspace $p=0$ of the phase space.
At least, one of the fixed point $x_{\Gamma}$ of the flow $\dot x = g(x)$ is an attractor.  Its basin of attraction may be limited by another fixed point $x_{\beta}$ which is unstable.
For the exponential map, the phase portraits of this Hamiltonian flow and the corresponding symplectic map are depicted in Fig. \ref{ppcdexp1}.

\subsection{The heteroclinic orbit and its action}

For the stochastic process (\ref{eqstoch2}), the attractor $x_{\Gamma}$ is found in a well of the kinetic potential defined by \cite{GT84,GT85}
\begin{eqnarray}\label{apot}
U(x) & = & -\int g(x) \, \mathrm{d}x \nonumber \\ 
&=& \frac{x^2}{2} - \int f(x) \, \mathrm{d}x
\end{eqnarray}
On the other hand, the unstable fixed point $x_{\beta}$ is the top of a barrier for the noise-induced escape from the potential well.  The escape over this barrier is an activated process with an escape rate given by Eq. (\ref{teedep}) where the constant $W_0$ is the action of a special trajectory of the Hamiltonian flow (\ref{flowsymp}).  This trajectory is given by the separatrix connecting the hyperbolic fixed point $\Gamma$ corresponding to the attractor to the other hyperbolic fixed point $\beta$ at the top of the barrier \cite{GT84,GT85,K88}.  See Fig. \ref{ppcdexp1}(a). The parametric equation of this separatrix is
\begin{equation}
p(x)=-2 \, g (x)
\end{equation}
and the action of this solution between the hyperbolic points $(x_{\Gamma},0)$ and $(x_{\beta},0)$  is given by
\begin{equation}\label{actcont}
W_{\rm h}^{\rm c} = \int_{x_{\Gamma}}^{x_{\beta}} p(x) \, \mathrm{d}x
\end{equation}
where the superscript `c' stands to recall that this is the action of the continuous-time Hamiltonian system.
This action yield the activation barrier for the escape problem in the stochastic process (\ref{eqstoch2}) from the attracting fixed point $x_\Gamma$ through the boundary $x_\beta$, as if the crossing by the point $x_\beta$ was the only possible channel of escape from $x_\Gamma$. 
The integration of the function $g$ gives the final result:
\begin{equation}\label{aacten}
W_{\rm h}^{\rm c}  = 2 \left[U(x_{\beta})-U(x_{\Gamma})\right]
\end{equation}
in terms of the kinetic potential (\ref{apot}).

\begin{figure}
\includegraphics[width=0.45\textwidth]{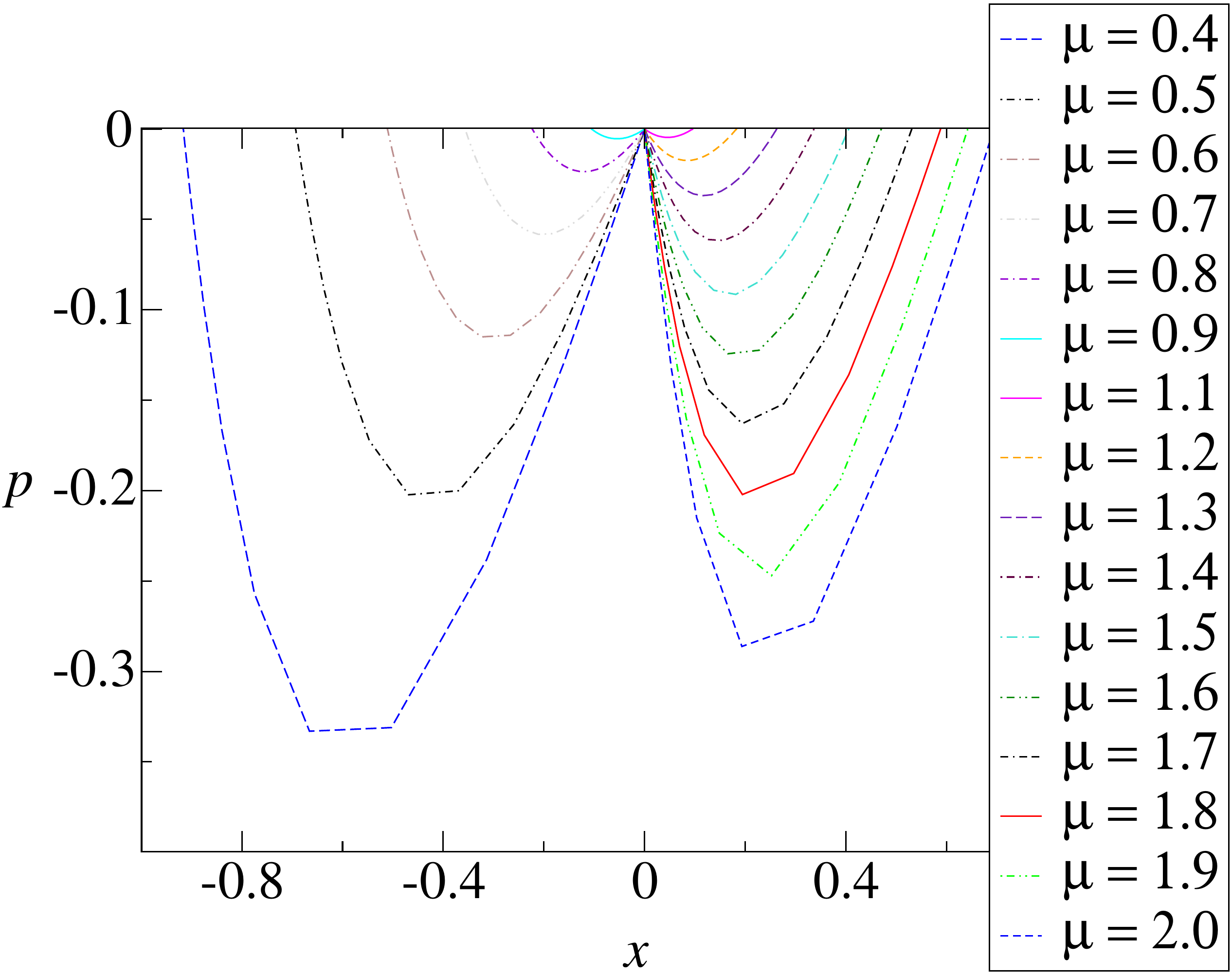}
\caption{\label{trajhetex} Evolution of the heteroclinic orbit of the symplectic map (\ref{mapsymp}) associated with the exponential map $x_{n+1} = \mu x_n \exp(-x_n)$ as the parameter $\mu$ varies from $0.4$ to $2$. Around the boundary bifurcation, the heteroclinic orbit tends to a continuous curve which is the separatrix of the Hamiltonian flow (\ref{flowsymp}).}
\end{figure}

We notice that the separatrix of the Hamiltonian flow (\ref{flowsymp}) is the limit of the heteroclinic orbit ${\rm h}={\cal W}^{\rm s}(\beta)\cap{\cal W}^{\rm u}(\Gamma)$ between the global stable set of the boundary point $\beta$ and the unstable manifold of the attractor $\Gamma$.  Therefore, we expect that the action (\ref{actcont}) or (\ref{aacten}) of the separatrix is the limit of the action of the heteroclinic orbit of the symplectic map evaluated by the sum:
\begin{equation}\label{h_fact}
W_{\rm h}= \frac{1}{2} \sum_{n=-\infty}^{+\infty} \left[x_{n+1} - f(x_n)\right]^2\Big\vert_{\rm h}
\end{equation}
The approximation of the map action (\ref{h_fact}) by the flow action (\ref{aacten}) is justified as the bifurcation is approached as seen in Fig.~\ref{trajhetex} where we observe that the heteroclinic orbit tends to the separatrix of the Hamiltonian flow (\ref{flowsymp}).

\section{Results \label{res}}

To investigate the relation between the \emph{activation barrier} $W_0$ and the actions (\ref{aacten}) or (\ref{h_fact}) for different types of bifurcations, we have carried out a Monte-Carlo simulation generating $2 \times 10^5$ escape events starting from the attractor $\Gamma$.  We have computed the distributions of escape times for different noise amplitudes $\varepsilon$. As these distributions slowly decrease exponentially in time in the limit $\varepsilon\to 0$, the best estimator of their average is simply the statistical average (\ref{esc_time}) according to Eq. (\ref{rate_time}).
In this way, we have obtained the mean escape time (\ref{esc_time}) as a function of $\varepsilon$. For $\varepsilon \to 0$, this function shows exponential increase according to Eq. (\ref{teedep}).
By fitting, we have computed the \emph{activation barrier} $W_0$. It is important to notice that Eq. (\ref{teedep}) is valid under the condition $\varepsilon << W_0$ on the values of the noise amplitude we had to consider.

In the case the one-dimensional map has several attractors, we have to consider the first-exit process from one basin of attraction as described in Subsection \ref{first-exit_time}.

\subsection{Transcritical bifurcation}

In this section, we study two simple maps which undergo a transcritical bifurcation.

\subsubsection{Logistic map}

We first consider the logistic map
\begin{equation}\label{logmap}
x_{n+1} = \mu \, x_n (1 - x_n)
\end{equation}
which is a well-known model in the fields of hydrodynamics \cite{K89,P91}, glass formation dynamics \cite{BR05}, and population dynamics \cite{N95,HS98}.
This map possesses two fixed points :
\begin{eqnarray}
x_1 & = & 0 \label{logistic_x1}\\
x_2 & = & 1-1/\mu \label{logistic_x2}
\end{eqnarray}
The logistic map undergoes a transcritical bifurcation at $\mu=1$ where these two fixed points exchange their stability. For $\mu < 1$, $x_1$ is stable while $x_2$ is unstable and forms with its non-trivial pre-image $f^{-1}(x_2)$ the boundaries of the basin of attraction of $x_1$: $B(x_1)=]x_2,f^{-1}(x_2)[$. For $\mu > 1$, $x_2$ is stable while $x_1$ is unstable and forms with its non-trivial pre-image the boundaries the basin of attraction of $x_2$: $B(x_2) = ]0,1[$. In both cases, the point at infinity is a second possible attractor for initial conditions outside the described basin of attraction.

\begin{figure}
\includegraphics[width=0.5\textwidth]{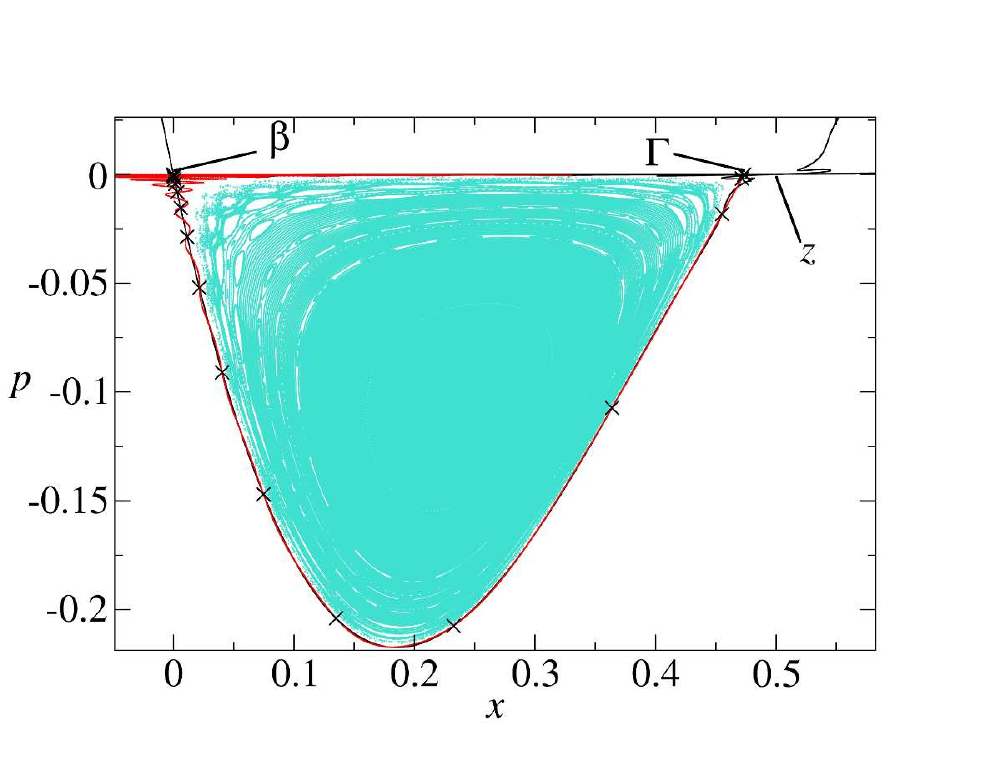}
\caption{\label{situ19} Phase portrait of the symplectic map (\ref{mapsymp}) associated with the logistic map (\ref{logmap}) for $\mu=1.9$.  Under this condition, the attractor corresponds to the fixed point $\Gamma=(x_2,0)$ with Eq. (\ref{logistic_x2}).  The fixed point of the escape barrier is $\beta=(0,0)$. Note that manifolds of the global stable set have to pass by the focal point $\boldsymbol{z}=(1/2,0)$ to cross the nondefinition set given by the singular line $\{(1/2,p): \, p \in \mathbb{R}\}$.  The heteroclinic orbit ${\rm h}={\cal W}^{\rm s}(\beta)\cap{\cal W}^{\rm u}(\Gamma)$ is depicted by crosses.}
\end{figure}

\begin{figure}
\includegraphics[width=0.45\textwidth]{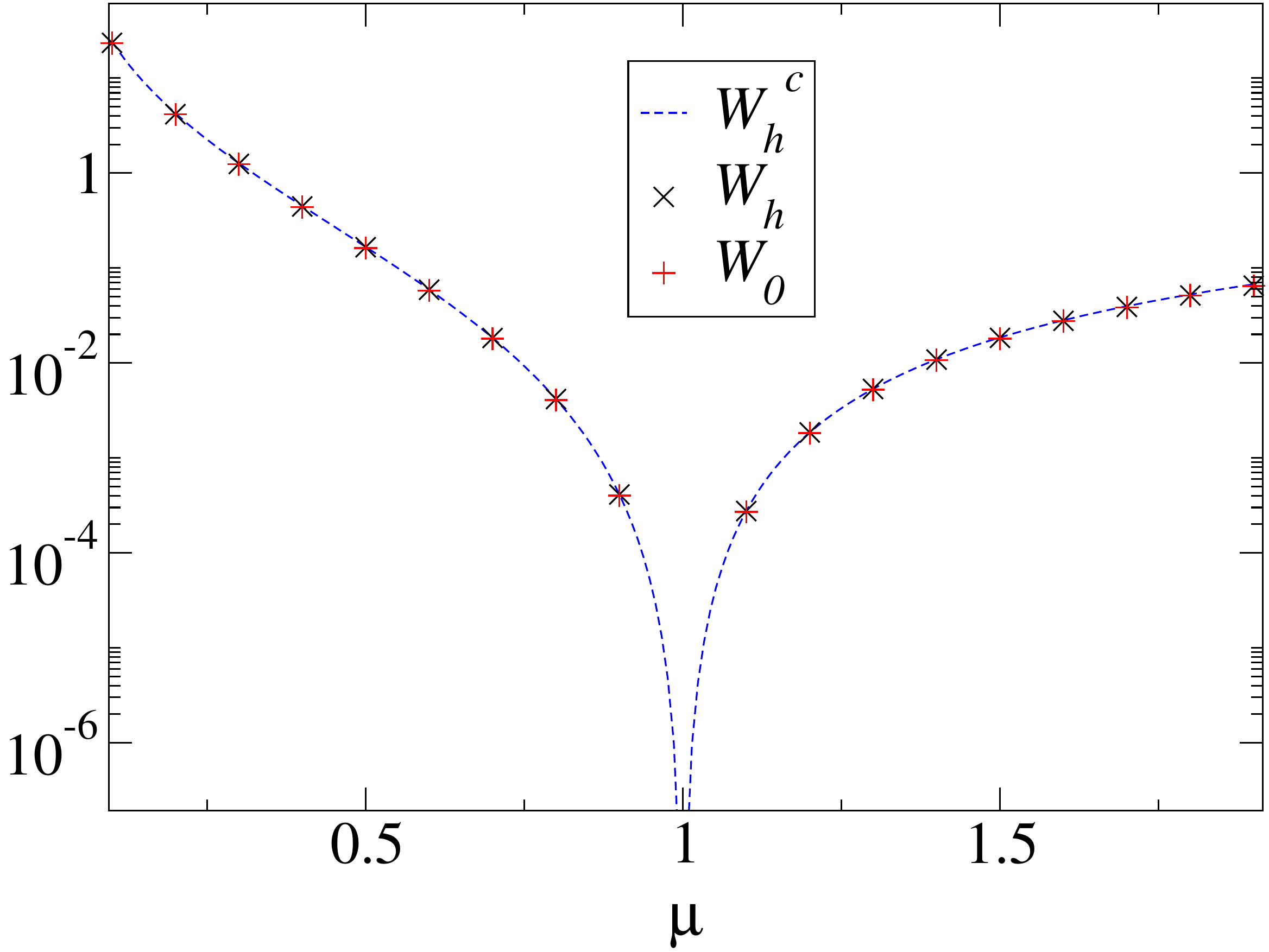}
\caption{\label{acthetlog} Action of the heteroclinic orbit $W_{\rm h}$ and activation barrier $W_0$ (with error bars) from the Monte-Carlo simulation versus the parameter value $\mu$ for the logistic map (\ref{logmap}). The blue hatched curve is the analytical action $W_{\rm h}^{\rm c}$ calculated near the transcritical bifurcation and given by Eq. (\ref{aelogi}).}
\end{figure}

For the noisy logistic map, the escape to infinity may occur via the two boundaries of the basin of attraction $B(\Gamma)$.  Both boundaries correspond to the same fixed point $x_{\beta}$ since one of the boundaries is the non-trivial pre-image $f^{-1}(x_{\beta})$ of the other. The phase portrait of the symplectic map (\ref{mapsymp}) is depicted in Fig.~\ref{situ19} where we observe similar structures as in Fig.~\ref{expps26} for the exponential map.  In particular, an elliptic island surrounds the non-trivial fixed point at $x=(\mu-1)/(2\mu)$ and $p=-(\mu-1)^2/(4\mu)$.  The symplectic map (\ref{mapsymp}) has two inverse maps $\phi_{\pm}^{-1}$, which are defined in the domain $p \geq x-(\mu/4)$ limited by the prefocal line (\ref{prefocal}). The map (\ref{mapsymp}) is not defined on the singular line $x=1/2$ where $f'(x)=0$ and a focal point exists at $\boldsymbol{z}=(1/2,0)$.  This focal point is the passage of the global stable set ${\cal W}^{\rm s}(\beta)$, as seen in Fig.~\ref{situ19}.  The heteroclinic orbit ${\rm h}={\cal W}^{\rm s}(\beta)\cap{\cal W}^{\rm u}(\Gamma)$ connecting the fixed points $\Gamma$ to $\beta$ was constructed with an algorithm described in Ref. \cite{Y98}. 
This heteroclinic orbit effectively controls the process of noise-induced escape to infinity of trajectories issued from the attractor.  Monte-Carlo simulations have allowed us to compute the escape rate and the activation barrier $W_0$, verifying in Fig.~\ref{acthetlog} that it coincides with the action of the heteroclinic orbit. Near the transcritical bifurcation, the continuous-time approximation provides us with the following analytical expression for this action:
\begin{equation}\label{aelogi}
W_{\rm h}^{\rm c} = \left| \frac{(\mu-1)^3}{3\mu^2} \right| 
\end{equation}
obtained from Eq. (\ref{aacten}) with the potential (\ref{apot}).
This formula applies to both sides of the transcritical bifurcation at $\mu=1$ because of the absolute value $\vert\cdot\vert$. Indeed, the fixed points (\ref{logistic_x1}) and (\ref{logistic_x2}) exchange their role of attractor and barrier at the bifurcation.  Figure~\ref{acthetlog} shows the excellent agreement between the activation barrier $W_0$ and Eq. (\ref{aelogi}) as the transcritical bifurcation is approached in the limit $\vert\mu\vert\to 1$.  At the bifurcation, the activation barrier vanishes as the cube of the control parameter $\Delta\mu=\mu-1$.  

\subsubsection{Exponential map} 

The exponential map
\begin{equation}\label{exmap}
x_{n+1} = \mu \, x_n \exp\left(-x_n\right) \qquad \mbox{with} \quad \mu >0
\end{equation}
shares several features with the logistic map.
It has two fixed points:
\begin{eqnarray}
x_1 & = & 0 \label{exp_x1}\\
x_2 & = & \ln\mu \label{exp_x2}
\end{eqnarray}
which cross each other in a transcritical bifurcation at $\mu=1$.
Moreover, the point at $x=-\infty$ is also attracting.
For $0<\mu < 1$, $x_1$ is stable, $x_2$ is unstable, and the basin of attraction of $x_1$ is $B(x_1) = ]x_2,+\infty[$. For $\mu > 1$, $x_2$ is stable, $x_1$ is unstable, and the basin of attraction of $x_2$ is $B(x_2) = ]x_1,+\infty[$.  Contrary to the logistic map, these basins of attraction are not compact and extend up to $x=+\infty$ where the exponential map vanishes and sends the points toward the fixed point $x_1=0$. The other half of the real axis is the basin of attraction of $x=-\infty$.

Phase portraits of the symplectic map (\ref{mapsymp}) are depicted in Figs.~\ref{expps26} and~\ref{exp1pp}, and discussed in Subsection \ref{phasespaces}.
The singular line is located at $x=1$ and the focal point at $\boldsymbol{z}=(1,0)$. According to Eq. (\ref{prefocal}), the prefocal line is $x=p+\mu/{\rm e}$.  The inverse (\ref{invmapsymp}) is not defined below this prefocal line for $p>x-\mu/{\rm e}$.

\begin{figure}
\includegraphics[width=0.45\textwidth]{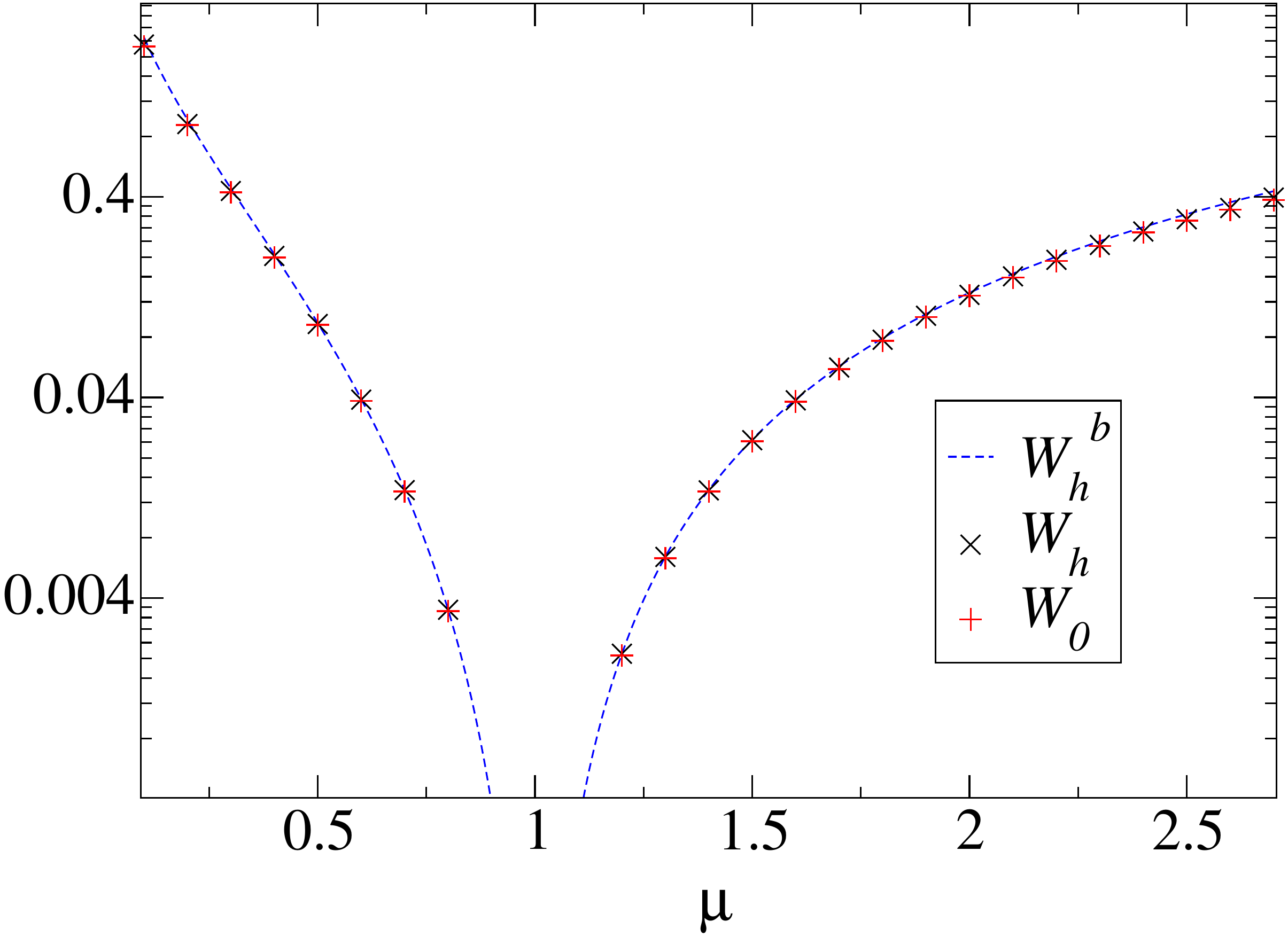}
\caption{\label{acthetex} Action of the heteroclinic orbit $W_{\rm h}$ and activation barrier $W_0$ (with error bars) from the Monte-Carlo simulation versus the parameter value $\mu$ for the exponential map (\ref{exmap}). The blue hatched curve is the analytical action $W_{\rm h}^{\rm c}$ calculated near the transcritical bifurcation and given by Eq. (\ref{aeexp1}).}
\end{figure}

An analytical calculation of the activation barrier can be carried out around the transcritical bifurcation at $\mu=1$ by using Eq. (\ref{aacten}) with the potential (\ref{apot}) yielding
\begin{equation}\label{aeexp1}
W_{\rm h}^{\rm c} = \left| 2 \mu - 2 - 2 \ln\mu - (\ln\mu)^2 \right|
\end{equation}
which applies below and above the bifurcation.  As for the logistic map, the activation barrier vanishes as the cube of the control parameter $\Delta\mu=\mu-1$.  
This result is verified in Fig.~\ref{acthetex} by comparison with the activation barrier computed by Monte-Carlo simulations.  A very good agreement is also observed with the action (\ref{h_fact}) of the heteroclinic orbit ${\rm h}={\cal W}^{\rm s}(\beta)\cap{\cal W}^{\rm u}(\Gamma)$.
It show evidence that the heteroclinic orbits play the key role in the determination of the activation barrier for the noise-induced escape from the attractor.

\subsection{Tangent bifurcation}

At the tangent bifurcation, a pair of stable and unstable fixed points emerge in the system.
A one-dimensional map illustrating this bifurcation is the quadratic map
\begin{equation}\label{quadmap}
x_{n+1} = \mu + x_n  - x_n^2
\end{equation}
which has the fixed points
\begin{equation}
x_{1,2} = \pm \sqrt{\mu}
\end{equation}
for $\mu \geq 0$. $x_1$ is stable while $x_2$ is unstable.
For $\mu>0$, the phase portrait of the symplectic map (\ref{mapsymp}) is similar to the one of the logistic map but the elliptic island no longer exists for $\mu <0$ where all the trajectories of the noiseless map {quadmap} escape to infinity.  For $\mu>0$, the escape form the basin of attraction of $x_1=x_{\Gamma}=+\sqrt{\mu}$ should be activated by the noise. Close to the tangent bifurcation at $\mu=0$, the critical slowing down justifies the treatment with the continuous-time approximation. Using 
Eq. (\ref{aacten}) with the potential (\ref{apot}), the activation barrier is here given by
\begin{equation}\label{aequadmap}
W_{\rm h}^{\rm c} = \frac{4}{3} \, \mu^{3/2}
\end{equation}
Here, we recover the result obtained by Beale that the exponent $3/2$ is universal for the activation barrier close to a tangent bifurcation~\cite{B89}.

We notice that the quadratic map (\ref{quadmap}) is transformed into the logistic map (\ref{logmap}) by the changes $x\to \mu x + (1-\mu)/2$ and $\mu\to(\mu-1)^2/4$.  This transformation is folding the parameter space two to one, therefore, mapping the transcritical bifurcation onto the tangent bifurcation.
In this way, the exponent $3$ of the transcritical bifurcation is consistent with the exponent $3/2$ of the tangent bifurcation \cite{B89}.

\subsection{Pitchfork bifurcation}

In the pitchfork bifurcation, a fixed point which pre-exists to the bifurcation destabilizes and generated two new stable fixed points.  These two new attractors are separated by the now unstable fixed point, which is the boundary between the two basins of attraction. At the bifurcation, the two attractors meet with this boundary so that the pitchfork bifurcation is a boundary bifurcation in this regard.

We are concerned by the first-exit process induced by the noise from one basin of attraction to the other across their common boundary, which constitutes a activation barrier between two wells for the kinetic potential (\ref{apot}).  As explained in Subsection \ref{first-exit_time}, this first-exit process is similar to an escape process in the sense that its rate is of Arrhenius type and also given by Eq. (\ref{teedep}) in the weak-noise limit.  At the pitchfork bifurcation, we expect that the activation barrier $W_0$ should vanish in a way characteristic of the bifurcation.  Two maps are investigated to determine this barrier.

\subsubsection{Cubic map}

The cubic map
\begin{equation}\label{cubmap}
x_{n+1} = \mu x_n (1 - x_n^2)
\end{equation}
undergoes a pitchfork bifurcation at the critical parameter value $\mu=1$
and presents bistability above this bifurcation for $\mu>1$.
For our purposes, we consider positive values of the parameter $\mu>0$.
The cubic map has one fixed point below the bifurcation and three above:
\begin{eqnarray}
x_1 & = & 0 \label{cubic_x1}\\
x_{2,3} & = & \pm\sqrt{1-1/\mu} \qquad\mbox{if} \quad \mu>1 \label{cubic_x23}
\end{eqnarray}
Moreover, the cubic map has an unstable period-two orbit at $x_{4,5}=\pm\sqrt{1+1/\mu}$.
For $\mu<1$,  the fixed point $x_1=0$ is an attractor and its basin of attraction is bordered by the period-two orbit: $B(x_1)=]x_5,x_4[$.  For $\mu>1$,  the fixed point $x_1=0$ becomes unstable and the fixed points $x_2$ and $x_3$ are two attractors with their basin of attraction given by $B(x_2)=]x_1=0,x_4[$ and $B(x_3)=]x_5,x_1=0[$, respectively.  Outside the interval delimited by the period-two orbit, trajectories can escape to infinity, which is a further attractor of the cubic map.  The symplectic map (\ref{mapsymp}) associated with the cubic map has three hyperbolic fixed points in the invariant subspace $p=0$ for $\mu>1$.  Now, there exist two elliptic islands symmetrically located around the origin. The origin $\beta=(0,0)$ stands for the barrier while $\Gamma=(x_2,0)$ for instance.  
A heteroclinic orbit ${\rm h}={\cal W}^{\rm s}(\beta)\cap{\cal W}^{\rm u}(\Gamma)$ connects the hyperbolic fixed point $\Gamma$ to $\beta$.  The associated symplectic map (\ref{mapsymp}) has two focal points $\boldsymbol{z}_{\pm}=(\pm 1/\sqrt{3},0)$ and two corresponding prefocal lines (\ref{prefocal}).

\begin{figure}
\includegraphics[width=0.45\textwidth]{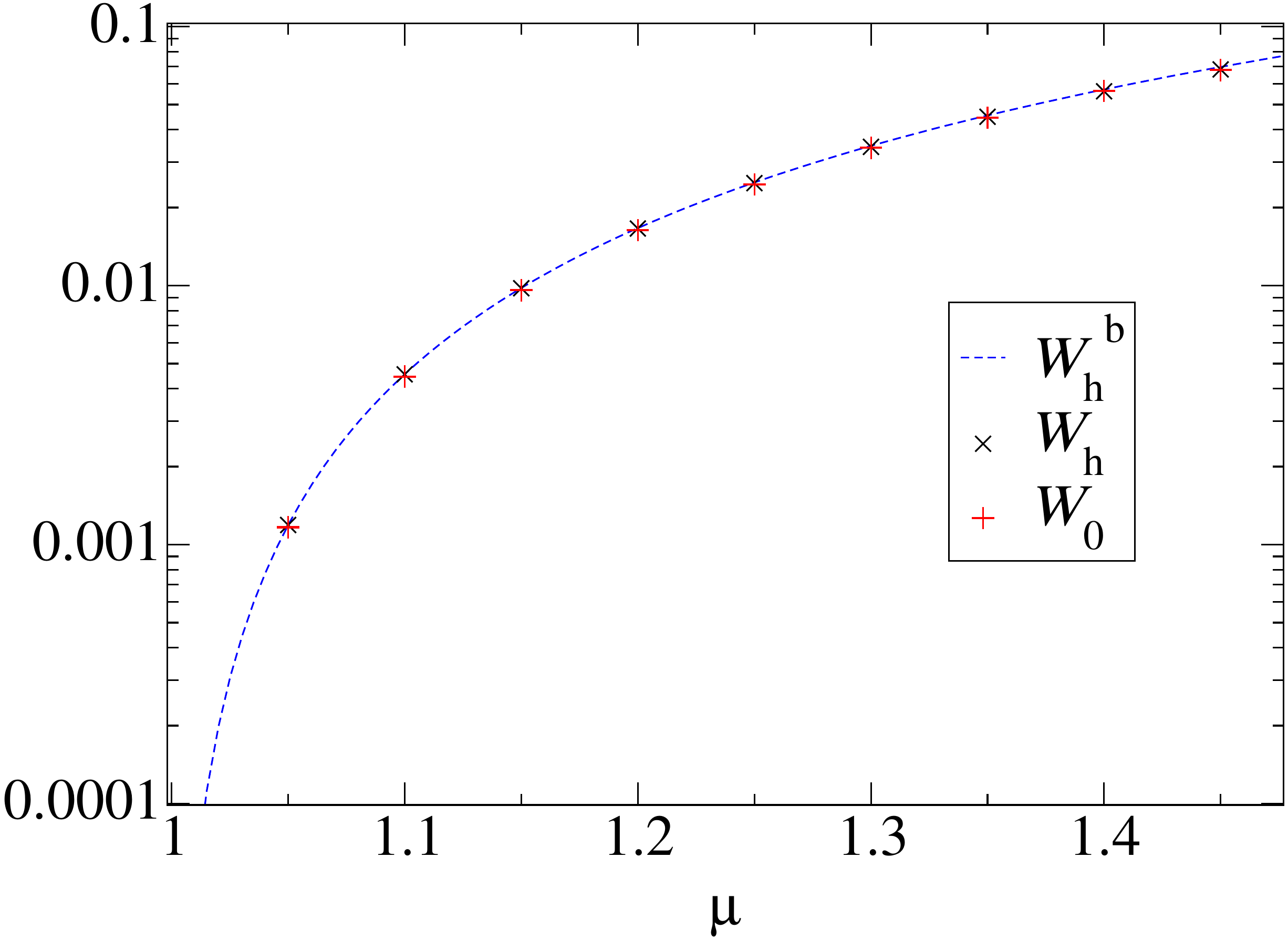}
\caption{\label{acthetcub} Action of the heteroclinic orbit $W_{\rm h}$ and activation barrier $W_0$ (with error bars) from the Monte-Carlo simulation versus the parameter value $\mu$ for the cubic map (\ref{cubmap}). The blue hatched curve is the analytical action $W_{\rm h}^{\rm c}$ calculated near the pitchfork bifurcation and given by Eq. (\ref{actancub}).}
\end{figure}

For $\mu>1$, noise induces random jumps between the two basins $B(x_2)$ and $B(x_3)$, as well as escape toward infinity.  The barrier between the two basins $B(x_2)$ and $B(x_3)$ is located at the origin $x_1=0$ and vanishes at the pitchfork bifurcation while the barrier for escaping toward infinity is the period-two orbit, which does not vanish around $\mu=1$.  Accordingly, the escape events remain rare and negligible with respect to the first-exit events from one basin to the other.  The difference of time scale between first-exit and escape events allows us to investigate the former without perturbation by the latter in Monte-Carlo simulations in the parametric domain of interest.  In the weak-noise limit, the first-exit process is controlled by the heteroclinic orbit ${\rm h}={\cal W}^{\rm s}(\beta)\cap{\cal W}^{\rm u}(\Gamma)$. Using the continuous-time approximation, we can calculate analytically the activation barrier near the pitchfork bifurcation with Eq. (\ref{aacten}) for the cubic map:
\begin{equation}\label{actancub}
W_{\rm h}^{\rm c} = \frac{1}{2 \mu} (\mu-1)^2
\end{equation}
which holds for $\mu>1$.  The excellent agreement with the results of Monte-Carlo simulations is shown in Fig.~\ref{acthetcub} where the action $W_{\rm h}$ of the heteroclinic orbit is also plotted.
Here, the activation barrier vanishes as the square of the control parameter $\Delta\mu=\mu-1$.

\subsubsection{Gaussian map}

Another example of bistable system is given by the Gaussian map:
\begin{equation}\label{exp2}
x_{n+1} = \mu \, x_n \exp(-x_n^2) \qquad \mbox{with} \quad \mu >0
\end{equation}
Its fixed points are
\begin{eqnarray}
x_1 & = & 0 \label{Gauss_x1}\\
x_{2,3} & = & \pm\sqrt{\ln\mu} \qquad\mbox{if} \quad \mu>1 \label{Gauss_x23}
\end{eqnarray}
The pitchfork bifurcation happens at $\mu=1$ and bistability manifests itself for $\mu>1$.
Contrary to the cubic map, escape to infinity is not possible for this map, even in the presence of noise.
In this regard, this model is very convenient to investigate the effects of noise on bistability.
For $0<\mu<1$, the whole real line is the basin of attraction of the origin $x_1=0$.
For $\mu>1$, the basins of attractions of the two attractors emerging at the pitchfork bifurcation are respectively $B(x_2)=]-\infty,0[$ and $B(x_3)=]0,+\infty[$.  Here also, the symplectic map (\ref{mapsymp}) has two focal points $\boldsymbol{z}_{\pm}=(\pm 1/\sqrt{2},0)$ and two corresponding prefocal lines (\ref{prefocal}).

\begin{figure}
\includegraphics[width=0.45\textwidth]{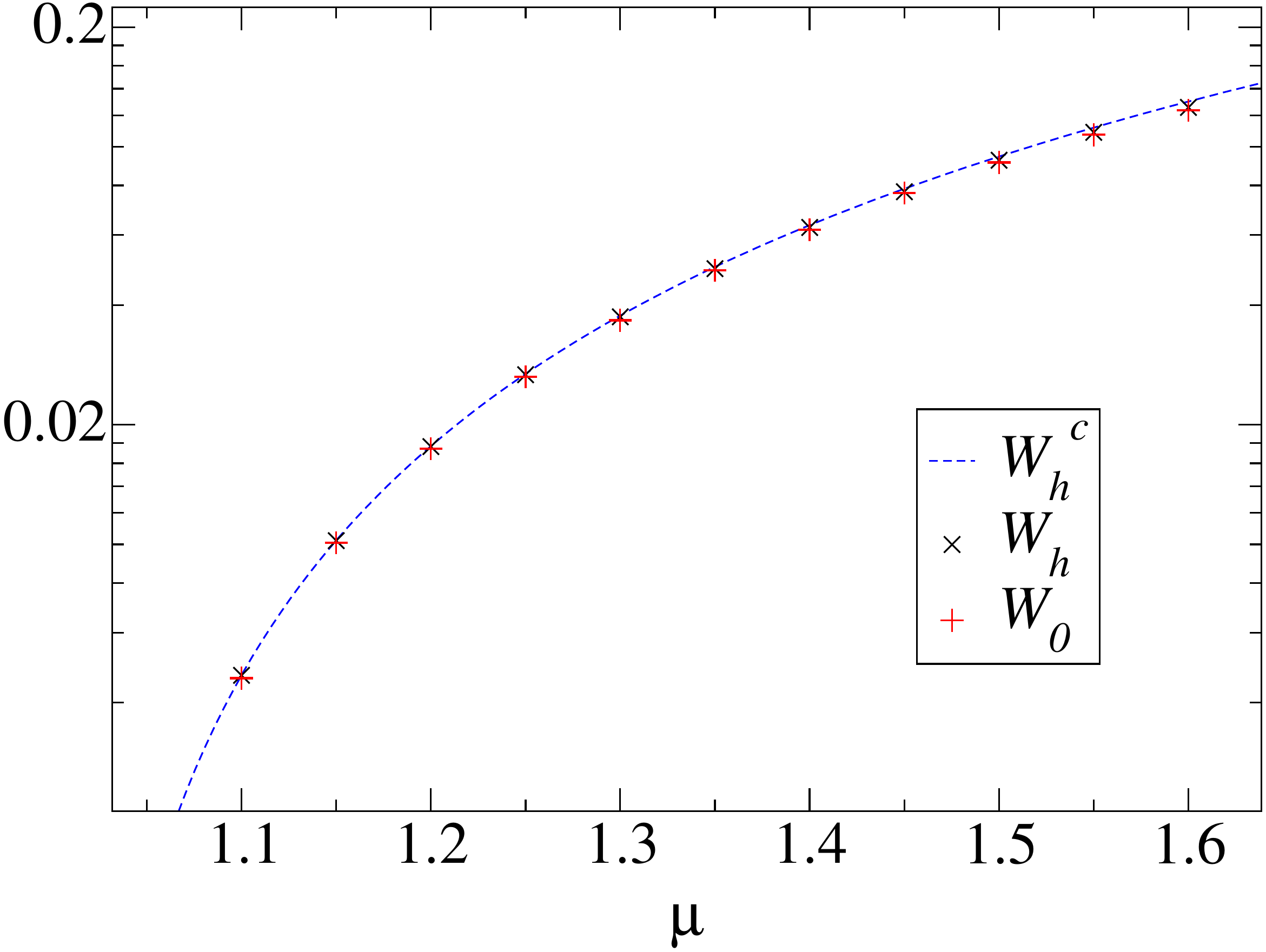}
\caption{\label{acthetex2} Action of the heteroclinic orbit $W_{\rm h}$ and activation barrier $W_0$ (with error bars) from the Monte-Carlo simulation versus the parameter value $\mu$ for the Gaussian map (\ref{exp2}). The blue hatched curve is the analytical action $W_{\rm h}^{\rm c}$ calculated near the pitchfork bifurcation and given by Eq. (\ref{actanGauss}).}
\end{figure}

Taking $x_\beta=0$ and $x_\Gamma=x_2$ for instance and calculating the action of the heteroclinic orbit ${\rm h}={\cal W}^{\rm s}(\beta)\cap{\cal W}^{\rm u}(\Gamma)$ in the continuous-time approximation, we find
\begin{equation}\label{actanGauss}
W_{\rm h}^{\rm c} = \mu -1 - \ln\mu
\end{equation}
which holds for $\mu>1$. Here again, the comparison with the results of Monte-Carlo simulations and the action of the symplectic map itself are excellent as shown in Fig.~\ref{acthetex2}.  As for the cubic map, the activation barrier vanishes as the square of the control parameter $\Delta\mu=\mu-1$, which characterizes the pitchfork bifurcation.

\section{Conclusions \label{conc}}

In this paper, we have studied noise-induced escape or exit from bifurcating fixed points of one-dimensional maps.  Such fixed points correspond to periodic orbits of strongly dissipative dynamical systems.  In this study, we have considered the transcritical, tangent, and pitchfork bifurcations \cite{N95}.  We followed a symplectic approach which applies in the weak-noise limit.  For noise generated by independent Gaussian random variables, the time evolution can be expressed in terms of path integrals defined by an action functional.  In the weak-noise limit, path integrals are dominated by the contributions of trajectories which are the extremals of the action functional.  For the present discrete-time systems, these trajectories are ruled by a symplectic map.  This two-dimensional area-preserving map reduces to the noiseless one-dimensional map on its one-dimensional phase space which is left invariant.  

The noninvertibility of the one-dimensional map and the associated symplectic map is at the origin of special phase-space structures.  In particular, the two-dimensional map is not defined on {\it singular lines}, except at their intersections with the invariant subspace.  These intersections are the so-called {\it focal points} \cite{BGM99,BGM03}.  The multiple inverses of the symplectic map are defined on domains bordered by so-called {\it prefocal lines} \cite{BGM99,BGM03}.  Strangely enough, the noninvertibility has for consequence that each focal point is the pre-image of a corresponding prefocal line.  Under such circumstances, invariant curves such as the stable and unstable sets may form loops attached to the focal points and they may be composed of several disjoint pieces, as here observed for the global stable set.  Otherwise, the symplectic map forms typical phase-space structures such as elliptic islands surrounded by chaotic zones and homoclinic or heteroclinic tangles.  Heteroclinic orbits have been identified which connect the hyperbolic fixed points corresponding to the attractor and the top of the barrier separating the basins of attraction.

Noise induces the escape or exit from each basin of attraction, changing the attractors of the noiseless map into metastable states.  The rates of escape or first exit have Arrhenius' dependence on the noise amplitude.  In the present non-thermal stochastic systems, the analogue of the activation energy is given by the action of the heteroclinic orbit connecting the attractor to the top of the barrier.  We have developed an analytical method to calculate these actions close to the bifurcations.  Thanks to the critical slowing down near the bifurcation, the noisy map behaves as a continuous-time stochastic system, whereupon the corresponding symplectic map can be approximated by a Hamiltonian flow with one degree of freedom \cite{OM53,FW84}.  For this flow, the action of the heteroclinic orbit can be calculated analytically.

\begin{table}
\begin{center}
\begin{tabular}{ccccccccccc}
\hline
\vline & bifurcation & \vline & action & \vline & exponent & \vline\\
\hline
\vline & transcritical & \vline & $\sim\vert\Delta\mu\vert^3$ & \vline  & $3$ & \vline  \\
\vline & tangent & \vline & $\sim\Delta\mu^{3/2}$ & \vline  & $3/2$ & \vline  \\
\vline & pitchfork & \vline & $\sim\Delta\mu^2$ & \vline  & $2$ & \vline  \\
\hline
\end{tabular}
\caption{Scaling behavior of the action $W_0$ of the activation barrier versus the control parameter $\Delta\mu\equiv\mu-\mu_{\rm c}$ and universal exponent for noise-induced escape or first exit from the attractor undergoing different bifurcations. The corresponding rate is given by $\gamma,\tilde\gamma \sim \exp(-W_0/\varepsilon)$ in terms of the noise amplitude $\varepsilon$.
\label{tab.bif}}
\end{center}
\end{table}

Applying this method to several one-dimensional noisy maps, we have shown how the activation barrier for noise-induced escape vanishes with the control parameter $\Delta\mu\equiv\mu-\mu_{\rm c}$ near the transcritical, tangent, and pitchfork bifurcations.  At these bifurcations, the activation barrier scales as $W_0\sim \vert\Delta\mu\vert^{\alpha}$ with a universal exponent $\alpha$ characteristic of the bifurcation, as summarized in Table \ref{tab.bif}.  The analytical results are in excellent agreement with Monte-Carlo simulations.

In conclusion, the symplectic approach is a powerful method to deal with stochastic dissipative systems in the weak-noise limit and to delineate the universality of their properties near bifurcations.

\begin{center}
$\ast \, \ast \, \ast$
\end{center}

The authors thank P. de Buyl, D. Andrieux and T. Gilbert for useful discussions and support.

\vspace{0.3cm}

{\bf Acknowledgments.}
This research is financially supported by the Belgian Federal Government 
(IAP project ``NOSY"), the ``Communaut\'e fran\c caise de Belgique''
(contract ``Actions de Recherche Concert\'ees'' No. 04/09-312), and
the F.R.S.-FNRS Belgium (contract F. R. F. C. No. 2.4577.04).


\end{document}